\documentclass[11pt]{article}
\usepackage[margin=1in]{geometry}
\usepackage{amsmath,amssymb,graphicx,booktabs}
\usepackage[hidelinks]{hyperref}
\usepackage[numbers,sort&compress]{natbib}
\newcommand{\bvec}{\mathbf{b}}
\usepackage{mathtools,amsthm}

\theoremstyle{definition}

\theoremstyle{remark}

\usepackage{xcolor}

\title{Burgers vectors from Voronoi topology}
\author{Emanuel A. Lazar\thanks{Department of Mathematics, Bar Ilan University, Ramat Gan 5290002, Israel, mLazar@math.biu.ac.il}}

\begin{document}
\maketitle

\begin{abstract}
Automated dislocation analysis is standard in atomistic simulations, yet current tools are limited to a handful of
crystal structures. We introduce an algorithm without adjustable parameters, based on Voronoi topology, that identifies dislocations
in a broad class of crystals. Each face shared by two cells that match the ideal crystal is taken as a step in it, and summing these steps around closed circuits of such cells yields exact Burgers vectors,
without geometric templates. Tested on
dislocations inserted into all 9{,}610 Materials Project structures whose Voronoi cells are stable with a wide
margin,
the algorithm finds every dislocation with its correct Burgers vector. In raw molecular-dynamics snapshots, including eleven dislocation types in $\theta$-Al$_2$Cu and bcc iron up to its
melting point, every Burgers vector it returns is correct, although near melting the dislocations merge into one
cluster. Voronoi topology thus provides a natural language in which a crystal, its
defects, and their Burgers vectors are all described.
\end{abstract}

\section*{Introduction}

Atomistic simulations have become a primary tool for studying dislocation nucleation, mobility, and interactions in
crystalline materials. Simulations of millions of atoms are now routine, and some reach hundreds of
millions~\cite{zepeda2021}, making automated dislocation identification essential. The Dislocation Extraction
Algorithm (DXA) developed by Stukowski and coworkers~\cite{stukowski2010dxa, stukowski2010defects, stukowski2012dxa,
stukowski2014jmps} established automated analysis as standard practice. DXA first classifies each atom using a
variant of common neighbor analysis~\cite{honeycutt1987, faken1994, stukowski2012cna}, comparing the pattern of bonds among its neighbors with those of ideal
reference structures to identify crystalline atoms. It then maps bonds between crystalline atoms onto ideal lattice
vectors through a structure template and sums these vectors around closed circuits, yielding the Burgers
vector~$\bvec$. Other approaches estimate~$\bvec$ from the Nye dislocation density tensor, computed from local
lattice distortions~\cite{hartley2005, dai2015}, or from slip vectors, the changes in the separations of neighboring atoms
relative to a reference configuration~\cite{zimmerman2001, hua2010}. A recent
graph-based method reconstructs dislocation loops and their Burgers-vector families from local displacement patterns in body-centered cubic (bcc),
face-centered cubic (fcc), and hexagonal close-packed (hcp) crystals~\cite{bhardwaj2026savi}.

These methods work for a limited set of crystal structures, each structure needing its own geometric template. DXA includes built-in templates for only five structures:
fcc, bcc, hcp, and cubic and hexagonal diamond~\cite{stukowski2012dxa, ovitodxa}. Extending the analysis to
additional structures requires dedicated effort. For example, adapting DXA to the C14 and C15 Laves phases required
modifying the neighbor analysis and restricting the analysis to a single sublattice~\cite{laca2021}. Other studies
rely on manual workarounds: running DXA only on the Sr and O sublattices of
SrTiO$_3$~\cite{okafor2025srtio3}, or, in ordered fcc Ni$_3$Al, using DXA for the Shockley partials and separately
highlighting the atoms of the antiphase boundaries, planar faults across which the ordered arrangement of the two species is
shifted~\cite{xu2024ni3al}. Studies of other complex crystals
analyze dislocations one sublattice at a time, as in MgSiO$_3$ perovskite~\cite{hirel2014}, or combine DXA with a tool written for that
structure, as in the CaAl$_2$ Laves phase~\cite{freund2026}. Consequently, many crystal
structures found in intermetallics, complex alloys, and minerals remain outside existing automated Burgers-vector tools,
although atomistic simulations increasingly target these
materials~\cite{zhang2024ceramics, zhou2017al2cu}.

The Burgers vector is a topological invariant: it depends only on how atoms are connected, not on their exact
positions, so in principle it can be computed from connectivity alone. The difficulty is deciding which atoms are connected. The usual choices, a fixed number of nearest neighbors or all atoms within
a cutoff distance, need a parameter tuned to each structure, and in configurations perturbed by strain or thermal
motion a single misassigned pair along a Burgers circuit can corrupt the result. Continuous measures of local order face a related limit: any continuous map from many coordinates to a few numbers gives the same value to some configurations that are arbitrarily far apart~\cite{landweber2016}, so no cutoff on such a value can separate structures cleanly. A discrete classification, constant on whole regions of configuration space, has no such limit. What is needed is a notion of adjacency
that needs no parameter, and whose robustness is a property of the crystal.

One such notion comes from the Voronoi tessellation, which divides space into cells, one per atom, each consisting of the
region closer to that atom than to any other~\cite{lazar2022ajp}. Two atoms are neighbors if their cells share a face. The \emph{topology} of a cell is the way its faces, edges,
and vertices fit together, regardless of their sizes and shapes: which faces meet along each edge, and which edges
meet at each vertex. The tessellation is built from the atomic positions, but its topology is discrete and
parameter-free~\cite{lazar2015pnas, vorotop2022}: it is typically unchanged by small displacements, and changes only when atoms
move far enough to create or remove a face or an edge~\cite{lazar2015pnas}. In an ideal crystal, the cells
of all atoms on one crystallographic (Wyckoff) site share the same topology, and in many crystals this topology is
stable against small perturbations, a property called \emph{topological stability}~\cite{votypes2026a, votypes2026b}. A recently released package, VoroTypes, computes the cell topologies of general crystal
structures, enumerating every topology that small perturbations can produce~\cite{votypes2026a,
votypes2026b}. For a topologically stable crystal, the only such topology is the ideal one, so its
robustness is known before any sample is analyzed.

Here we show that Burgers vectors can be read from Voronoi topology in crystals whose cells are topologically
stable. Each face shared by two cells whose topologies match
those of the ideal crystal is read as the corresponding step in the ideal crystal. Atoms whose cells do not match form defect clusters, and closed loops
through the layer of crystalline atoms around each cluster give its Burgers vector. The only input is the ideal crystal structure, and in these crystals the result is exact. We validate the method on all 9{,}610
Materials Project structures whose cells are stable with a wide margin and on raw molecular-dynamics snapshots up to
the melting point of iron, and apply it to complex intermetallics under active plastic deformation.
Crystals whose cells are not topologically stable are more complicated, and they include the most common metals,
fcc and hcp. We show that the same ideas handle them in practice, with weaker guarantees, and that they extend to
$\omega$-titanium, a degenerate crystal for which DXA has no template. A general treatment of such crystals is
left for future work.

\section*{Results}

\subsection*{Burgers circuits in two dimensions}
\begin{figure}[!htbp]
\centering
\includegraphics[width=\textwidth]{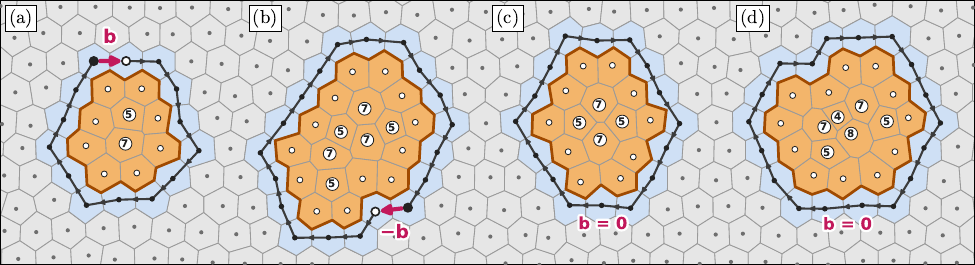}
\caption{\textbf{Defect clusters in a triangular crystal with thermal noise.} Dots represent atoms and polygons their
Voronoi cells; non-hexagons show their number of sides. Each defect cluster (orange) is separated from the crystal by
its boundary (dark orange outline); the crystalline cells along it form the shell (blue), through which the circuit (black) runs
clockwise from the open circle to the filled one. (a) A dislocation: the circuit fails to close, and the red step
that closes it is the Burgers vector $\bvec$. (b) A dislocation of opposite sign whose core has merged with a thermal fluctuation; the
circuit returns $-\bvec$. (c) A bond flip, in which two neighboring atoms move apart and two of their common
neighbors move together; $\bvec=0$. (d) Two adjacent bond flips. The cluster looks much like that in (b), but its circuit returns $\bvec=0$.}
\label{fig:concept}
\end{figure}

Although our method targets three-dimensional crystals, its ideas are most clearly introduced in two dimensions; we
return to three dimensions in the next subsection. In two dimensions, a dislocation is a point defect
(Fig.~\ref{fig:concept}a). Far from the defect, the crystal appears defect-free, yet it cannot be mapped continuously
onto an ideal crystal everywhere: traversing a closed path around the defect reveals a relative displacement defined by
the Burgers vector~$\bvec$. The standard method to measure $\bvec$ is the Burgers circuit (Fig.~\ref{fig:concept}).
One steps along neighboring atoms around the defect until returning to the starting atom, recording each step as its
corresponding vector in the ideal crystal. In defect-free crystal, these step vectors sum to zero; around a
dislocation, they yield a nonzero sum equal to $\bvec$. Equivalently, if one step is left out so that the others would
sum to zero, the walk around the dislocation fails to close, and the step needed to close it is $\bvec$
(Fig.~\ref{fig:concept}a). We traverse circuits clockwise as seen with the dislocation line, normal to the plane, pointing toward the viewer.
Summed from start to finish, the steps of such a circuit give the same $\bvec$ as the finish-to-start, right-hand
(FS/RH) convention~\cite{hirthlothe}, whose circuit runs the opposite way.

Our method rests on three topological properties of the Burgers vector. First, it is the same for every circuit that
encloses the same dislocations and passes only through defect-free crystal, whatever the circuit's size or shape.
Second, it vanishes for circuits that enclose no dislocation. Third, the vectors of several enclosed dislocations add.

In two dimensions, automating the Burgers circuit requires two decisions for atoms displaced from their ideal
positions: which atoms are neighbors, and the ideal direction of each step. In a triangular crystal, Voronoi topology
provides both the neighbors and the directions~\cite{lazar2024twod}. Two atoms are neighbors if their
Voronoi cells share a side. A polygon contains little information, so the topology of an atom in two dimensions is
taken to include the numbers of sides of its neighbors, in order around the cell~\cite{lazar2024twod}. Unlike in three
dimensions, it therefore depends on the neighboring cells as well as on the cell itself. Away from defects,
nearly all cells remain hexagons even at high temperatures, while a dislocation core appears as a pentagon beside a
heptagon~\cite{nelson1979} (Fig.~\ref{fig:concept}a). Each hexagon has its own \emph{frame}, a numbering of its sides by the six directions of the crystal. The side two
hexagons share fixes how
their frames are related (Supplementary Fig.~\ref{fig:si-frames2d}a); turning every step of a loop of hexagons into a
common frame and summing the ideal steps gives $\bvec$.

In a two-dimensional triangular crystal, an atom is classified as crystalline if its cell and the cells of all its neighbors are hexagons~\cite{lazar2024twod}.
Non-crystalline atoms are grouped into connected \emph{defect clusters}. Each cluster is separated from the surrounding
crystal by its \emph{boundary}, the closed curve of Voronoi sides between the cluster's cells and crystalline ones, and
the crystalline cells along the boundary form the cluster's \emph{shell} (Fig.~\ref{fig:concept}a). Following the
boundary once around, stepping through the shell cells beside it, gives the Burgers circuit, which returns the
cluster's net Burgers vector. A cluster without holes has a single boundary, and so a single circuit: nothing is searched for or chosen. Small defect clusters also appear at finite temperature, such as the
pentagon--heptagon--pentagon--heptagon cluster created by a single bond flip; their circuits return $\bvec=0$, so
they are not counted as dislocations (Fig.~\ref{fig:concept}c). The circuit does not depend on what the cluster
looks like: a dislocation core merged with a thermal fluctuation still returns its Burgers vector (Fig.~\ref{fig:concept}b; Supplementary Movie~1),
while a cluster of fluctuations that appears disordered returns zero (Fig.~\ref{fig:concept}d).

Most of this framework extends directly to general crystal structures and to three dimensions: two atoms are neighbors
if their Voronoi cells share a face, each shared face relates the frames of the two cells, and Burgers vectors are
evaluated along the boundary between each defect cluster and the surrounding crystal, through the shell of crystalline
cells beside it. Two aspects require generalization. First, whereas the triangular crystal has a single type of cell,
with frames related by rotations of a hexagon, atoms on distinct Wyckoff sites generally have different cells; each
site therefore requires its own reference cell, and frames are related by the symmetries of the crystal. Second, in
three dimensions the boundary of a defect cluster is a surface rather than a curve, and a surface with several handles
can enclose a dislocation network whose branches have different Burgers vectors.

\subsection*{Three dimensions: defects, shells, and Burgers vectors}
In three dimensions, a dislocation is a line, and its type depends on the angle between the line and $\bvec$: edge
if they are perpendicular, screw if parallel, and mixed at any angle in between. For a perfect dislocation, $\bvec$ is
a \emph{lattice vector}, a translation that maps the ideal crystal onto itself. Perfect dislocations may dissociate into partial dislocations, whose
Burgers vectors are not lattice vectors, joined by a ribbon of stacking fault, a planar defect that interrupts the
stacking sequence of layers.

We now take the two changes in turn. First, the reference. For each Wyckoff site of the ideal crystal, the Voronoi
cell of one atom serves as a reference cell, and each of its faces is labeled by the ideal vector to the neighbor
across it~\cite{votypes2026a}. A sample cell whose topology matches that of a reference cell receives these labels
through a one-to-one correspondence of faces, edges, and vertices that preserves how they meet (a combinatorial
isomorphism). We call this correspondence a \emph{labeling}, and it gives the cell its local frame. A labeling assigns ideal positions to a cell's atom and its neighbors. Two neighboring cells therefore both assign
positions to themselves, to each other, and to their common neighbors. If one symmetry of the crystal maps the first
cell's positions of these atoms onto the second cell's, the two labelings \emph{agree}, and this symmetry is the cells' \emph{frame relation}. A cell is \emph{crystalline}, as in earlier
work, if its topology matches that of a reference cell~\cite{lazar2015pnas, vorotop2022}. A Burgers circuit needs
slightly more: each step from one crystalline cell to the next needs a frame relation, which exists only if the two
cells' labelings can agree. A crystalline cell with no labeling that agrees with those of its crystalline neighbors,
for example because they are not arranged around it as in the crystal, cannot be joined to them by frame relations.
For the purpose of constructing Burgers circuits, it is therefore treated as a defect.
Composing frame relations around a closed loop of crystalline cells gives a net rotation and a net translation. Around
a dislocation the net rotation is the identity, and the net translation is the Burgers vector, with the FS/RH
convention above. A loop \emph{closes} if its net rotation is the identity and its net translation is zero. In topologically stable
crystals, where every Voronoi vertex is shared by exactly four cells, as in bcc metals, B2 alloys, and Laves and
$\theta$ phases, every loop of crystalline cells that can be shrunk to a point through crystalline cells closes
(Supplementary Section~\ref{sec:si-exact}). Any two loops that can be deformed
into one another through crystalline cells therefore give the same Burgers vector.

Second, the boundary. As in two dimensions, non-crystalline atoms are grouped, now through shared faces, into defect
clusters. The boundary of a cluster, the Voronoi faces between its cells and crystalline ones, is a closed surface, and
the crystalline cells along it form the cluster's shell, one atom thick (Supplementary Fig.~\ref{fig:si-shell}). The boundary is still unique, but a surface has many closed curves, so the circuit is no longer
determined by the cluster alone. What matters is how a curve winds around the surface, which depends on its genus $h$, its number of handles. The boundary need not be connected, and for
each of its connected pieces, $h$ follows from Euler's formula, $V-E+F=2-2h$, where $V$, $E$, and $F$ are the numbers of
vertices, edges, and faces of the surface. A spherical boundary ($h=0$), such as those around point defects
and thermal fluctuations, returns zero, since every curve on it can be shrunk to a point. A dislocation
line or loop is encased by a torus ($h=1$), on which a curve once around the tube returns $\bvec$ and a curve along the tube that does not wind around it
returns zero. For a straight line through the periodic box, a curve along the tube closes only through the box's
periodic boundary, so its net translation is not a Burgers vector, and it is not used (Methods). A torus can also
return zero, for example around a ring-shaped thermal cluster. Surfaces of higher genus arise, for example, around
dislocation networks, and a thermal fluctuation can add a handle to any cluster.

Each curve on the boundary defines a walk through the shell cells beside it. The shell therefore contains a
circuit for every curve on the boundary, and no thicker layer is needed. A curve's vector depends only on how it winds
around the boundary surface, and only a few curves are needed: standard methods for surfaces give exactly two
independent curves per handle~\cite{eppstein2003} (Methods). The resulting net translations are the
cluster's Burgers vectors. At a junction Burgers vector is conserved, the vector of one branch being the sum
of those of the others, because a curve around one branch can be deformed continuously on the surface into the combination of curves around the others.

This construction contains no adjustable parameters. Voronoi topology identifies the defect atoms, combinatorics
determines the clusters, their boundary surfaces, and the curves on them, and exact frame relations yield the Burgers
vectors. The vectors are exact under two conditions on the crystal: (1) it is topologically stable, and (2) it is not
\emph{topologically pseudo-symmetric}, that is, its Voronoi tessellation is not combinatorially more symmetric than the crystal.
Crystals that fail the first are \emph{degenerate}: some Voronoi vertices are shared by more than four cells, and the
topology of the cells around them changes under arbitrarily small perturbations, as in fcc and hcp metals. We treat
them below, with weaker guarantees. The simplest example of the second is an oblique two-dimensional lattice, whose
cells are all hexagons, as in the more symmetric triangular lattice (Supplementary Fig.~\ref{fig:si-oblique}). In such crystals topology alone can give more
than one inequivalent Burgers vector, and the method chooses by geometry (Methods).

\subsection*{Validation}
We validated the method on samples with known dislocations, inserted into ideal crystals. Some samples were analyzed
directly, after small random displacements of the atoms; others were first evolved by molecular dynamics (MD) under
conditions in which the dislocations neither annihilated nor combined. In both cases the true Burgers vectors are the
inserted ones. Neighboring dislocations may still share one cluster, when the non-crystalline regions around them
touch, and such a cluster returns their vectors and their sums. A test counts as wrong if any vector found is none of
these.

\emph{The Materials Project.} We quantify how far a crystal's Voronoi topology is from degenerate by its \emph{Voronoi stability gap} $g$, which VoroTypes reports as the minimum relative gap~\cite{votypes2026a, votypes2026b}. In three
dimensions, each Voronoi vertex is the center of a sphere that passes through
four or more atoms and contains none inside (in two dimensions, Supplementary Fig.~\ref{fig:si-gap}). In a degenerate crystal, some
sphere passes through more than four atoms, and $g=0$. Otherwise, at every vertex, the next nearest atom lies at least
a fraction $g$ farther away than the four on the sphere, so atoms must move by a distance of order $g$ times the
nearest-neighbor distance before any cell can change its topology. We computed $g$ for all 163{,}211 structures of the
Materials Project~\cite{jain2013mp}. Most (71\%) have $g>0$ (above $10^{-6}$, the rounding of the stored coordinates),
but their gaps are mostly tiny, with a median of $g=0.003$; the other 29\% are degenerate. A subset of 9{,}610
structures (5.9\%) has $g\ge0.2$ (Supplementary Table~\ref{tab:mp-stats}), so every Voronoi cell is stable by a
wide margin, comparable to the stable crystals we test by MD below ($g=0.61$ for bcc and B2, $0.50$ for C15, and $0.29$
for $\theta$-Al$_2$Cu at the potential's lattice parameters). These 9{,}610 structures span all seven crystal systems
and 103 space groups, have 1 to 57 Wyckoff sites, and fall into 58 families of cell topologies (distinct sets of reference-cell topologies) and 242 distinct
geometries (structures alike up to scale when species are ignored); 2{,}302 are experimentally observed. About two thirds of them (6{,}353, in 29 geometries) are ordered
compounds whose atoms, ignoring species, form a bcc arrangement, such as Heusler and CsCl-type phases; the other
3{,}257, in 97 space groups and 213 geometries, include 1{,}463 experimentally observed structures and such
intermetallic types as Laves phases (C15, C14, C36), A15, CaCu$_5$, $\theta$-Al$_2$Cu (C16), and the $\sigma$ and
$\mu$ phases. The full list, with each structure's identifier, gap, and results, is
available with the data (Data availability).

\begin{figure}[!htbp]
\centering
\includegraphics[width=0.965\textwidth]{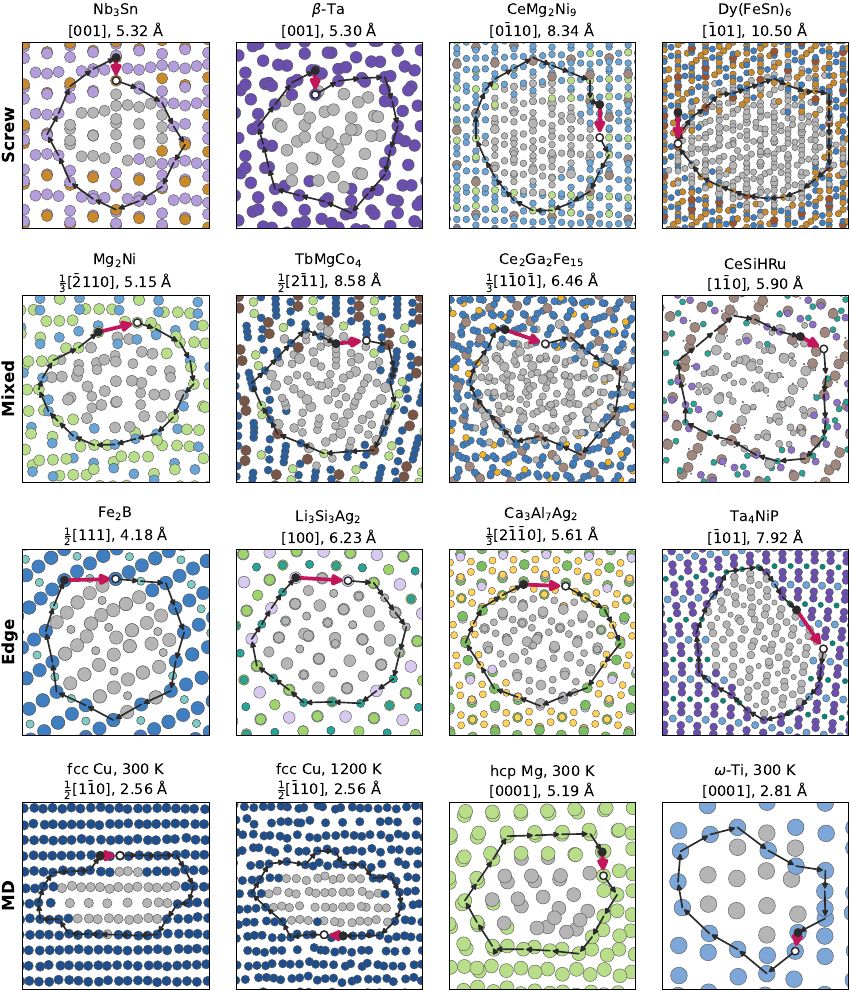}
\caption{\textbf{Burgers circuits.} Rows 1--3: crystals of the test set, with screw, mixed ($30$--$60^\circ$), and edge dislocations chosen as in
Supplementary Section~\ref{sec:si-gallery}, not the test's own dipoles.
Row 4: raw MD snapshots of edge dislocations in fcc Cu, where the defect cluster holds both Shockley partials and the
stacking fault, and of screw dislocations in hcp Mg and $\omega$-Ti. Each panel gives $\bvec$ in the conventional cell
and its length. Edge dislocations are viewed along the line, the others $30^\circ$ off it. Atoms of the defect cluster
are gray, and the others are colored by element. From the open circle to the filled one, the black steps run clockwise
and would close in an ideal crystal. The red arrow that closes them is $\bvec$, up to the strain of the crystal.}
\label{fig:gallery}
\end{figure}

Each structure was first snapped onto exact positions of its declared space group (Methods). We then inserted
dislocation dipoles (pairs of dislocations with opposite Burgers vectors), each into its own fully periodic supercell (Methods). Two dipoles have their lines along the shortest lattice vector $\mathbf s$: a screw dipole, with
$\bvec=\mathbf s$, and a non-screw dipole, whose Burgers vector is the shortest lattice vector not parallel to $\mathbf
s$; the non-screw dipole is pure edge in 1{,}775 structures (18\%) and mixed, at $60$--$89^\circ$ to the line, in the
rest. In 9{,}270 structures, each dipole gave two separate clusters, with $+\bvec$ and $-\bvec$, in supercells about
$28d$ wide ($d$ the nearest-neighbor distance; 2{,}028--12{,}312 atoms). The remaining 340 structures did so in
supercells two or four times as wide. These have smaller gaps (median $g=0.23$, against $0.61$ for all 9{,}610), so
noise and the strain around a dislocation change cell topologies farther from its core than in crystals with larger gaps
(Discussion), and wider samples are
needed. A third dipole, with $\bvec=\mathbf s$ and its line along the shortest lattice vector not parallel to $\mathbf s$, gave the
same picture: 9{,}468 structures at $28d$ and 142 in wider samples. Across all 29{,}360 analyses, no vector was wrong (Fig.~\ref{fig:gallery}; Supplementary Section~\ref{sec:si-val}). DXA ignores chemical species~\cite{ovitodxa}, so it can be applied to the bcc-based
structures with its bcc template; in 20 of them chosen at random, it also correctly returned the inserted vectors of all 40
dipoles, two per structure. The other 3{,}257 lie outside its templates. Smaller gaps are
discussed below.

\emph{Molecular dynamics.} All MD samples are fully periodic supercells containing either four straight dislocations with
zero net Burgers vector (a quadrupole) or a single prismatic dislocation loop in bcc Fe (Methods). Without quenching or time
averaging, we analyzed 846 raw snapshots of seven crystals (Supplementary Table~\ref{tab:md}). In bcc Fe they contain edge and
screw $\tfrac12\langle111\rangle$ dislocations and a prismatic vacancy loop, from 300 to 1755~K (up to
$0.99\,T_\mathrm m$, with $T_\mathrm m=1770$~K the melting point of the potential~\cite{mendelev2003}, Methods).
The others contain $[111]$ screw \emph{superdislocations} in B2 NiAl at 300 and 900~K, eleven
combinations of Burgers vector and line in $\theta$-Al$_2$Cu at 300 and 500~K, and screw, edge, and $60^\circ$ mixed
$\tfrac12\langle110\rangle$ dislocations in C15 MgCu$_2$ at 300~K. A superdislocation has a Burgers vector that
preserves chemical order, here $[111]$, twice the bcc vector $\tfrac12[111]$. In crystals with degenerate cells (next
section), they contain edge dislocations in fcc Cu from 300 to 1200~K and $[0001]$ screw dislocations in hcp Mg and
$\omega$-Ti at 300 and 600~K.

No snapshot returned a wrong vector. In 703 of the 846, each dislocation forms its own cluster, which returns
its inserted vector. In the other 143, the non-crystalline regions around neighboring dislocations touch, so these dislocations share one
defect cluster, which returns their vectors and their sums. Shared clusters occurred in 15 of the 110 $\theta$ snapshots at 500~K, all for the two largest Burgers vectors, $[110]$
($|\bvec|=8.4$~\AA) and $[100]$, and in bcc Fe
near melting: at $0.85\,T_\mathrm m$ in 2 of 42 snapshots, and at 0.93, 0.96, and $0.99\,T_\mathrm m$ in all 126. Near $T_\mathrm m$, thermal fluctuations leave a quarter to a third of the atoms classified as non-crystalline, and their clusters percolate; the box-spanning cluster returns only multiples of $\bvec$, from the dislocations it contains. In fcc Cu each edge dislocation dissociates into two
Shockley partials, which, together with the stacking fault between them, form one cluster with the Burgers vector of the perfect dislocation,
$\tfrac12[1\bar10]$ (next section); every Cu snapshot from 300 to 1200~K gave four such clusters. In Mg and
$\omega$-Ti every snapshot gave four separate clusters with $\pm[0001]$, and in B2 NiAl four with $\pm[111]$. Analyzed instead with the species-blind bcc reference (next section), the same NiAl snapshots return the $\tfrac12[111]$ vector of the
\emph{superpartials}, the two $\tfrac12[111]$ dislocations into which a superdislocation dissociates, although in no snapshot were all of them in separate clusters. None was wrong.

We also ran DXA~\cite{stukowski2012dxa,ovito} on the bcc Fe edge and screw samples from 300 to 1650~K
($0.93\,T_\mathrm m$). In every snapshot it found segments with the inserted $\pm\tfrac12[111]$ and no
other vector. At $0.93\,T_\mathrm m$ it still traced separate line segments, although for the edge dislocations only
about 60\% of their length (Supplementary Section~\ref{sec:si-val}). DXA has no structure template for $\theta$, C15, or $\omega$-Ti.

\subsection*{Degenerate cells and partial dislocations}
Not every crystal is topologically stable, and the most common metals are not. A general treatment of such crystals is
beyond this paper. Here we show two things. First, a treatment built for fcc and hcp metals. Second, a route not specific to one structure, which we apply to $\omega$-titanium.

In fcc and hcp metals, the ideal Voronoi cells are degenerate, as several of their vertices are shared by more than
four cells. We call the twelve atoms whose cells share a face with an atom's cell in the ideal crystal its \emph{full
neighbors}, and the six that touch it only at a degenerate vertex its \emph{partial neighbors}~\cite{votypes2026a}.
Thermal fluctuations split the degenerate vertices randomly, so no cell keeps the ideal topology, and most
gain small faces shared with partial neighbors, which a cell's topology alone cannot always tell apart from faces
shared with full neighbors. Because each degenerate vertex can split in only finitely many ways, the topologies that
small perturbations can produce can be enumerated exhaustively: 2{,}815 for fcc and 9{,}556 for
hcp~\cite{lazar2015pnas, votypes2026b}. A given topology may arise from the ideal cell in more than one way, each with its own set of
twelve faces belonging to full neighbors; we call each such set a \emph{topological alignment} of the cell with the ideal one, or simply an
\emph{alignment}. The enumeration gives a table of every perturbed topology and its alignments. An alignment plays the part that a labeling plays in stable crystals, but fixes only which faces
join full neighbors; the ideal vectors are assigned when these neighbors are matched to their ideal arrangement
(Methods).

Evaluating cells individually leaves a choice of alignment for 10--12\% of the atoms in fcc Cu and 62--64\% in hcp Mg in
our MD snapshots. Neighboring cells, however, must agree: a shared face joins full neighbors for both cells or for
neither. Enforcing this face-by-face agreement reduces the ambiguous cells to at most 0.06\% in Cu and 0.6\%
in Mg (Methods). Away from defects, agreement never removes the true alignment, since the
true alignments of neighboring cells always agree. Near a defect core, however, a single remaining alignment can be
wrong. A cell is therefore treated as a defect unless its twelve neighbors form their ideal arrangement and its frame
relations with the crystalline cells beside it are crystal symmetries. Cells that fail these checks, or are left with
several alignments or none, join the defect cluster. With 94--98\% of the Cu atoms and 96--97\% of the Mg atoms passing these checks, the method finds
every dislocation in these samples with its correct Burgers vector (Fig.~\ref{fig:gallery}; Supplementary Section~\ref{sec:si-val}).

Stacking-fault atoms have the full-neighbor arrangement of the other close-packed structure (hcp-type in fcc,
fcc-type in hcp). Most of them fail these checks and join the defect cluster, so a dissociated dislocation and its
stacking fault form a single cluster with the Burgers vector of the perfect dislocation. In fcc Cu, the
$\tfrac16[2\bar1\bar1]$ and $\tfrac16[1\bar21]$ Shockley partials of an edge dislocation give one cluster with
$\tfrac12[1\bar10]$. Some atoms, 12--17\% of the stacking-fault atoms and 16--37\% of those at the partial cores, pass the checks
with a wrong full neighbor, but these wrong alignments agree with
one another and with the crystal around them, so every smallest loop still closes and every vector was correct
(Supplementary Section~\ref{sec:si-degenerate}). DXA, by contrast, recognizes stacking-fault atoms by pattern matching
and traces partial dislocations individually~\cite{stukowski2012dxa}. Resolving partials in the present framework is
left for future work.

Crystals with stable cells have their own counterpart of partial dislocations, which can be resolved by choosing
the reference. In B2 NiAl, a $[111]$ superdislocation dissociates into two $\tfrac12[111]$ superpartials joined by an
antiphase boundary. The reference may include the species of the atoms or ignore them (Methods). With species, in
the ordered B2 reference, a cell is treated as a defect if its neighbors, with their species, are not arranged around
it as in B2. The cells along the antiphase boundary are then defects, and the two superpartials and the boundary form
one cluster with $[111]$. Without species, in the bcc reference, the antiphase boundary is defect-free bcc, and a
circuit can enclose a single superpartial and return its $\tfrac12[111]$, although in our snapshots the
superpartials were never all separated. The B2 reference cannot do this, since every circuit
through defect-free B2 returns a lattice vector of B2, and $\tfrac12[111]$, which joins a Ni site to an Al site, is
not one.

The same enumeration works beyond close-packed metals. In the $\omega$ phase of titanium (P6/mmm, three atoms per
unit cell), an atom on the Wyckoff 2d site has 11 full neighbors and six partial neighbors. Thermal motion opens each of the six
contacts into a face or removes it, so at finite temperature its cell has 11 to 17 faces. Enumerating every combination gives 382
cell topologies at this site and 8 at the other, 388 in all, since two occur at both~\cite{votypes2026a}; these serve
directly as reference cells (Methods). In defect-free samples with noise of up to $0.05\,d$ per component,
95--98\% of the atoms are crystalline, and no cluster returns a vector. In all 27 dipoles inserted as in
the Materials Project test, with $\bvec=\mathbf c$ or $\bvec=\mathbf a$, both dislocations were found with their
correct vectors, as were all four lines in each of the 38 MD snapshots (Fig.~\ref{fig:gallery}; Supplementary Section~\ref{sec:si-degenerate}). DXA has no template for
this crystal.

\subsection*{Complex crystals in action}
\begin{figure}[!htbp]
\centering
\includegraphics[width=\textwidth]{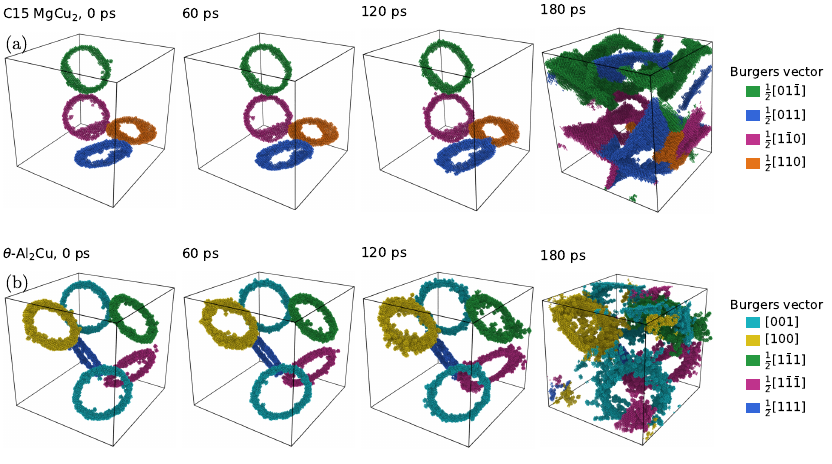}
\caption{\textbf{Dislocations in complex intermetallics during plastic flow.} (a) C15 MgCu$_2$ and (b)
$\theta$-Al$_2$Cu at 0, 60, 120, and 180~ps, from periodic samples of $1.2\times10^{6}$ atoms strained at 300~K.
Only defect clusters with a nonzero Burgers vector are drawn, each atom colored by the vector of its branch (key at
right).}
\label{fig:showcase}
\end{figure}

As an application, we simulated plastic deformation of two complex intermetallics: $\theta$-Al$_2$Cu, which lies
outside the templates of existing automated tools, and C15 MgCu$_2$, for which the only automated analysis we know of is a tool written specifically for Laves
phases~\cite{laca2021}. Each sample contains $1.2\times10^{6}$ atoms at 300~K in a
fully periodic box, with seeded dislocation loops (Methods). Plastic flow is driven by straining
the box at a constant rate, and the raw snapshots are analyzed directly (Fig.~\ref{fig:showcase}); times are counted from the
start of loading. Beyond the seeded loops, the evolution is not known in advance, so we rely on two consistency checks:
each seeded loop must keep its vector until it meets another defect, and after merging, every branch of the resulting
network must have the vector of the loop it came from or, where loops react, a sum of such vectors. In
Fig.~\ref{fig:showcase}, each core atom (atom of a defect cluster) is colored by the Burgers vector of its branch, read from the shortest
circuits through the shell near it (Methods).

In C15 MgCu$_2$, a Laves phase with two Wyckoff sites, four of the six seeded loops survive
relaxation (minimization and 10~ps at 300~K before loading). For the first
151~ps each is a defect cluster of its own, whose boundary is a torus with a single Burgers vector; the four are
distinct $\tfrac12\langle110\rangle$ vectors of the seeded slip systems (Fig.~\ref{fig:showcase}a). The loops then
grow, and at 152~ps two of them meet: they form one cluster, on whose surface curves return the vectors of both branches and their sum, and by 167~ps all four form a single network of $6.7\times10^4$ atoms (Fig.~\ref{fig:showcase}a), which
keeps growing. In all 181 snapshots, taken every 1~ps, the circuit around every branch returns one of the four
seeded vectors. Applied with default parameters to the Mg sublattice, which forms a diamond lattice, as in
LaCA~\cite{laca2021}, DXA finds all four vectors until the loops meet and three or four of
them afterward, but it then traces only a fraction of the line. It also reports segments with other vectors, among
them $\tfrac16\langle112\rangle$, which may be synchro-Shockley partials~\cite{xie2023synchroshockley} that our analysis
does not resolve (Supplementary Section~\ref{sec:si-dxa}). Before the loops meet, DXA's lines follow our clusters but leave gaps and a few isolated
segments with open ends.
By about 180~ps the moving dislocations have left planar faults behind them. As with
stacking faults in fcc metals, the faults join the defect clusters, which still return the exact total vectors of the
dislocations they contain. Resolving the partial dislocations that bound the faults would require recognizing the atoms of the faults as crystal of their
own structure, with frame relations to the host, so we follow C15 only up to this point (Supplementary Movie~2).

We follow $\theta$-Al$_2$Cu for 200~ps (Supplementary Movie~3). Of its six seeded loops, three have
$\tfrac12\langle111\rangle$ vectors, one has $[100]$, and two have $[001]$, on different planes
(Fig.~\ref{fig:showcase}b). For about
100~ps they remain six separate tori, apart from frequent brief contacts, in which two loops touch, share one cluster
with the vectors of both and their sum, and separate again. They then merge, at first parting again from time to time. At 150~ps two
clusters remain, and from 180~ps on all six loops form a single network (Fig.~\ref{fig:showcase}b). Throughout, no branch with a new
vector forms. In all 1{,}001
snapshots, taken every 0.2~ps, the circuit around every branch returns one of the seeded vectors.

\begin{figure}[!htbp]
\centering
\includegraphics[width=\textwidth]{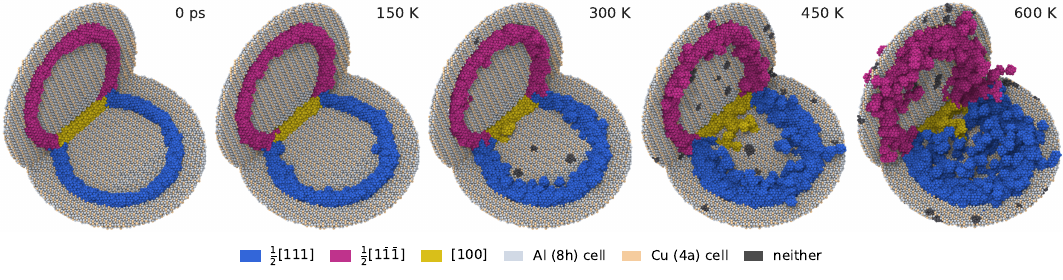}
\caption{\textbf{Two loops in $\theta$-Al$_2$Cu joined along a segment with $\tfrac12[111]+\tfrac12[1\bar1\bar1]=[100]$.} Two loops
whose dislocation lines cross, after energy minimization (0~ps) and at the end of holds at 150, 300, 450, and 600~K. Atoms of
the defect cluster are colored by the Burgers vector of their branch, and crystal atoms on the two glide planes by the
cell topology of their site (dark: neither ideal topology).}
\label{fig:junction}
\end{figure}

To follow two loops whose vectors add, we placed two $\tfrac12\langle111\rangle$ loops on intersecting $\{211\}$ planes, with vectors
$\tfrac12[111]$ and $\tfrac12[1\bar1\bar1]$ that add to $[100]$ (Methods). Where the two dislocation lines cross, the loops join
during relaxation, and a circuit around the joined segment returns $[100]$. Circuits alone cannot tell whether the lines
combine there into one core. The segment persists
at 300~K throughout a 359~ps run (Supplementary Movie~4). A copy held at 150~K and then heated in steps to 600~K keeps the $[100]$ segment in every
snapshot, while its loops roughen but do not shrink (Fig.~\ref{fig:junction}; Supplementary Movie~5). In a second sample, each loop stops 8~\AA\ short of the
line where the two planes meet. The loops soon touch, and a segment with Burgers vector $[100]$ forms where they
touch, comes apart, and forms again. After about 80~ps it is present almost continuously (Supplementary Movie~6). In
a third, with each loop 11~\AA\ short of that line, no $[100]$ segment formed within 50~ps.

\section*{Discussion}
We have shown that Burgers vectors can be derived from the Voronoi topology of an atomic configuration. In
topologically stable crystals without pseudo-symmetry the result is exact, with no adjustable parameter. In practice,
the method needs only the ideal crystal structure, as given in a crystallographic file, from which everything it
uses is derived. Across the 9{,}610
structures tested, it returned no wrong vector and found every inserted dislocation separately, some only in wider
samples. A third of them, in 97 space groups across all seven crystal systems, lie outside DXA's templates; among
them are Laves phases, A15 compounds, $\sigma$ and $\mu$ phases, and CaCu$_5$- and $\theta$-Al$_2$Cu-type structures.

Three further differences from template methods follow from the topology. First, because circuits run on
the boundary surface of a defect cluster, whatever its size, no maximum circuit length is needed. DXA, in contrast,
does not find dislocations whose cores are too wide for its longest trial circuit~\cite{ovitodxa}. Second, the core needs no separate criterion: we take it to be the dislocation's defect cluster, the atoms whose
cells are not crystalline or are treated as defects. DXA defines the core by a separate criterion~\cite{ovitodxa}. Nor does the core's structure need to be recognized: two clusters that look alike can
differ in what they hold (Fig.~\ref{fig:concept}b,d), since that is decided by the crystal around them. Third, the
output always conserves Burgers vector. Because every defect cluster is bounded by a closed surface and every vector
comes from a closed curve on it, no dislocation can end inside crystal, and Burgers vector is conserved at every junction by
construction. DXA's traced lines can have open ends, as in the C15 sample above (Supplementary
Fig.~\ref{fig:si-c15-dxa}a). DXA does, however, do three things our method does not. It traces
lines, with their directions and lengths, whereas we report the Burgers vectors of clusters and their atoms. It traces the two partial
dislocations of a dissociated dislocation in fcc and hcp metals separately, and near melting it still traces separate
lines where our clusters have merged (Supplementary Section~\ref{sec:si-val}). The second could be added to our approach, in which
stacking-fault atoms are recognized as crystal of the other stacking, with frame relations between the two, and we plan to
address it in future work. The third reflects what the two methods report, as discussed
below.

Degenerate crystals are handled with weaker guarantees. For fcc and hcp metals we built a specific treatment, in which
each cell's alignment is fixed by agreement between neighbors and a check of its twelve full neighbors. For
$\omega$-titanium, every topology that small perturbations can produce becomes a reference cell. VoroTypes enumerates these topologies automatically, without code
specific to the structure, so we expect this route to extend to many, perhaps most, of the degenerate and small-gap
crystals, though this remains to be tested. In both, exactness is no longer guaranteed near defects. Instead, the software checks that
the smallest circuits there return zero, and every vector in our tests was correct. For fcc and hcp metals the method
is built for thermal snapshots, in which thermal motion splits the degenerate vertices randomly, and does not apply to
energy-minimized samples. A general treatment, tested beyond
these three crystals, is left for future work.

Beyond degenerate cells, the method has three limitations. First, our guarantees are stated for fully periodic samples, the usual setting of
bulk simulations, in which every defect cluster is enclosed by a closed surface in the crystal. Where a free surface
cuts the crystal, it is no longer clear what a Burgers circuit is when part of it would lie outside the sample; we
leave this case to future work. Second, the method requires the ideal cell topology to survive around the defects.
Dislocation strain, about $|\bvec|/(2\pi r)$ at distance $r$ from the line, destroys it where it exceeds $g$, within a
distance of order $|\bvec|/(2\pi g)$, so in crystals with small gaps $g$
the topological core, the region of cells classified as non-crystalline, extends far beyond the core itself, and
clusters grow and merge rather than return wrong vectors. This affects many crystals: 71\% of Materials Project
structures have $g>0$, but in most of them the gap is so small that thermal vibrations destroy the ideal cell
topology. We have therefore restricted our study of stable-cell
crystals to $g\ge0.2$. Treating small-gap crystals as
degenerate, by enumerating perturbed cell topologies and enforcing neighbor agreement, is one possible remedy, though it
requires a table for each structure; we leave it, with a general treatment of degenerate cells, to a separate study.
That study has a wide scope: about 93\% of Materials Project structures match no fcc, hcp, bcc, or diamond arrangement even when
species are ignored, most of them with small positive gaps. Third, in topologically pseudo-symmetric crystals the
Voronoi tessellation is more symmetric than the crystal, as in any oblique two-dimensional lattice, whose cells are
all hexagons (Supplementary Fig.~\ref{fig:si-oblique}). Topology alone then leaves one discrete choice of labeling per region of crystalline cells, which the
method makes from the fit of the whole region's geometry, reporting its margin (Methods). Pseudo-symmetry is a
property of the ideal crystal, can be detected before any sample is analyzed, and affects at most about 4\% of the
structures we tested (Supplementary Section~\ref{sec:si-alg}).

The method needs no adjustable parameters because the three ingredients of a Burgers circuit come from a single object, the
Voronoi tessellation. In particular, a circuit needs to know which atoms are neighbors, how their local frames are related, and which
smallest loops must close. Template methods obtain these by separate procedures. In our
framework, Voronoi neighbors share a face, so adjacency is mutual. Both cells take their frame relation from that same
face, so it is consistent in both directions. And the smallest loops, around Voronoi edges, provably close in stable
crystals (Supplementary Section~\ref{sec:si-exact}). In such crystals the three cannot disagree.

At finite temperature, ``crystalline'' and ``defect'' have no unique definition, and existing tools
supply operational ones through their criteria: the neighbor patterns accepted by common neighbor
analysis~\cite{honeycutt1987, faken1994} and polyhedral template matching~\cite{larsen2016ptm}, the maximum circuit
length of DXA~\cite{stukowski2012dxa}, or cutoffs on distances and order parameters. The Burgers vectors are
fixed by how a sample was made, and every method should return them. But which atoms a method counts as core, and whether nearby
dislocations appear as one cluster or as separate lines, depend on its definitions, and on these points methods can be compared only for consistency.
Near melting this distinction matters. In bcc Fe at 0.93--0.99\,$T_\mathrm m$, Voronoi topology classifies a quarter to a
third of the atoms as not crystalline, and these regions join the dislocations into one
cluster. The common neighbor analysis in DXA finds even more, 46\% at $0.93\,T_\mathrm m$, yet DXA still traces
separate lines there. Voronoi topology gives a definition without adjustable parameters, under which every quantity we computed
is consistent: the Burgers vectors are exact, and clusters that enclose no dislocation return zero.

Conceptually, the construction can be seen as a discrete version of the classical geometric theory of defects, in which the
crystal is treated as a continuum whose lattice directions define a local frame at every point, and a dislocation
appears where steps along these directions fail to close around a loop~\cite{bilby1955, kleman2008}. Here the frames are the cells' labelings, and the relations between
neighboring frames are obtained exactly from Voronoi topology. Because these relations include rotations as well as translations, the construction measures more than Burgers vectors: a net rotation around a loop would mark a
disclination.

Because the method analyzes raw thermal configurations directly, without minimization or time averaging, it can be
applied snapshot by snapshot to entire molecular-dynamics trajectories. Each analysis also partitions the sample into
connected regions of crystalline cells, each with its own crystal frame, and the non-crystalline cells can themselves
be classified by their Voronoi topology, which distinguishes point defects and core structures~\cite{lazar2015pnas, vorotop2022}. Voronoi topology thus offers a single description of crystal structure,
defects, and their Burgers vectors.

\section*{Methods}

\subsection*{Overview}
The analysis takes the atomic positions, the periodic box, and the ideal crystal structure, and proceeds in four steps. (1) The
Voronoi tessellation of the sample is computed. (2) Each atom is classified as crystalline if the topology of its Voronoi cell matches that of a reference cell, and as a defect atom otherwise. Further, a crystalline atom that cannot be joined to its crystalline neighbors by frame relations is treated as a defect atom for the Burgers analysis (below). (3) Defect
atoms whose cells share a face are grouped into defect clusters, and the boundary of each cluster is extracted as a
closed surface, with its genus. (4) Curves on each boundary, two per handle (none on a sphere), are evaluated through the shell as compositions of
frame relations, giving the cluster's Burgers vectors. Each step works cell by cell or cluster by cluster, so the work grows in proportion to the number of atoms, except when one cluster holds much of the sample (below).

\subsection*{Voronoi tessellation}
Voronoi tessellations are computed with voro++~\cite{voro}, using its multithreaded extension~\cite{lu2023voromt}.
For every cell we record its faces as oriented cycles of vertices, and the atom across each face. A Voronoi vertex
is identified by the set of cells that meet there, so two cells sharing a face agree on its vertices without
reference to coordinates.

\subsection*{Reference cells and labelings}
The reference is built from the ideal crystal structure alone: its unit-cell vectors, fractional coordinates, and,
optionally, species.
For each Wyckoff site (an orbit of the space group, found with spglib~\cite{spglib}) the Voronoi cell of one atom serves
as a reference. It stores the ideal vector to the neighbor across each face and the ideal position of each vertex.
Labelings are found as combinatorial isomorphisms onto the reference cell of the same topology~\cite{weinberg1966, lazar2012prl}.
When a cell's topology has more symmetries than its site has in the crystal, it has several labelings. All are kept,
and those incompatible with the labelings of its neighbors are pruned by arc consistency~\cite{mackworth1977}, a
constraint-propagation method that repeatedly removes each labeling with which no remaining labeling of some neighbor agrees. A cell left with
no labeling cannot be joined to some neighbor by a frame relation, so it is treated as a defect, and propagation restarts without it. The result does not depend on the order in which cells are processed (Supplementary Section~\ref{sec:si-alg}). Neighbors usually rule such extra labelings out, because their own cells do not
share the extra symmetry. Only when the whole tessellation shares it, under topological pseudo-symmetry (Results), does a
choice remain that topology cannot make (below). If species are given, each atom must have its site's species.

\subsection*{Frame relations}
The frame relation of two labeled cells $i$ and $j$ that share a face is the crystal symmetry $(R,\mathbf d)$,
with $R$ in the point group $\mathcal G$ of the crystal and $\mathbf d$ the ideal vector from $i$ to $j$, that maps the ideal
positions of the face's vertices given by one cell's labeling onto those given by the other's:
\begin{equation}
\mathbf P_i(v_k)=R\,\mathbf P_j(v_k)+\mathbf d,\qquad k=1,\dots,m ,
\label{eq:gluing}
\end{equation}
where $\mathbf P_i(v)$ is the ideal position that the labeling of cell $i$ gives vertex $v$. Only ideal positions
enter, so the frame relation is fixed by the two labelings, and the sample's coordinates play no part (its
computation, including a tolerance for rounding of the input structure, is described in Supplementary
Section~\ref{sec:si-alg}). Loops never cross a face whose two labelings no crystal symmetry relates in
this way. In a two-dimensional triangular crystal, the frame relation reduces to a rotation fixed by the numbers
that two hexagons assign to their shared side (Supplementary Fig.~\ref{fig:si-frames2d}a).

\subsection*{Choosing labelings}
Where a cell still has several labelings, they are chosen by growing regions of crystalline cells from a seed, the
unlabeled cell with the fewest remaining labelings. The
seed's labeling fixes the region's frame: a different choice would rotate every frame of the region alike and leave
every Burgers vector unchanged. A neighboring cell is labeled once topology decides it, that is, once all of its
labelings consistent with its labeled neighbors are equivalent (related by symmetries of its site). These differ only by a symmetry of the cell's site, which changes no
Burgers vector, so we take the first. A cell whose consistent labelings
are not all equivalent waits until more of its neighbors are labeled. Regions are grown through all crystalline cells,
not only those around defect clusters, because agreement may have to spread from farther away to settle a cell near a defect, and no fixed distance would always suffice.

Without pseudo-symmetry, this labels a large defect-free region correctly, since the only labeling that all its cells
accept is the true one, up to the frame. In practice, choices remain open only near defects, where information is
missing. If every remaining cell is waiting, the one with the most labeled neighbors is decided by geometry,
taking the labeling with the smallest misfit (below), and the software counts and reports such decisions. In the defect-free samples of all 9{,}610 structures of the
Materials Project test, no such choice occurred, except in one structure that cannot be snapped onto its declared space
group (Supplementary Section~\ref{sec:si-alg}).

Under topological
pseudo-symmetry the seed itself has several inequivalent labelings, which topology cannot tell apart. The region is
then grown from each in turn, and the labeling whose cells fit best overall is kept, with its margin over the
runner-up reported (Supplementary Section~\ref{sec:si-alg}, with a two-dimensional example). The fit is measured by the misfit: the
root-mean-square deviation of the actual neighbor vectors from the ideal ones after the best orthogonal map and
uniform scaling. Of the defect-free samples of the 9{,}610 structures in the Materials Project test, each snapped onto
its declared space group where possible, 390 (4.1\%) have several such labelings (an upper bound, Supplementary Section~\ref{sec:si-alg}). In 73 of
them the runner-up's misfit exceeds the best by less than 0.1~\AA, and in 3 by less than 0.01~\AA. All 9{,}610 were
nevertheless analyzed correctly (Results).

\subsection*{Loops and Burgers vectors}
Composing the frame relations around a closed loop of cells $i_0\to i_1\to\dots\to i_n=i_0$ gives a net rotation
$R$ and a net translation $\mathbf t$, $(R,\mathbf t)=f_{i_0i_1}f_{i_1i_2}\cdots f_{i_{n-1}i_0}$, with
$f_{ij}=(R_{ij},\mathbf d_{ij})$ the frame relation of Eq.~\eqref{eq:gluing}, composed as
$(R_1,\mathbf d_1)(R_2,\mathbf d_2)=(R_1R_2,\,R_1\mathbf d_2+\mathbf d_1)$. Around a dislocation the net rotation
is the identity (no rotation), and for a loop that runs clockwise when the line points toward the viewer, $\mathbf t$
is the Burgers vector $\bvec$ of the FS/RH convention (Results). The software checks
that every smallest loop of crystalline cells in the sample, around a single Voronoi edge, closes. In topologically
stable crystals this is guaranteed (Supplementary Section~\ref{sec:si-exact}), so a loop that does not close signals a problem with the input, such as a periodic box too
short along a dislocation line, and is reported.
A loop's Burgers vector then depends only on the defects it encloses, and it is automatically a lattice
vector. For topologically stable crystals it is the true Burgers vector whenever the loop can be deformed through
crystalline cells into crystal far from all defects, provided the labeling agrees with the true
crystal at one cell. Topology guarantees this agreement except under topological pseudo-symmetry, where the
region-level fit decides it (above). Rotations are transported along frame relations, so all Burgers vectors are expressed in one
crystal frame per crystalline region (connected region of crystalline cells), and vectors around different clusters
can be compared directly. Evaluated in separate frames, two dislocations with the same Burgers vector could return it
as $\tfrac12[111]$ and as $\tfrac12[\bar111]$, and a junction would not visibly conserve it. No region is singled out as
the bulk. Each connected piece of a boundary surface faces a single crystalline region in stable crystals, and its
curves are evaluated in that region's frame; vectors are reported together with their region.

\subsection*{Defect clusters, boundary surfaces, and loops}
\emph{Clusters and boundary.} A defect cluster $D$ is a connected component, through shared Voronoi faces, of the
atoms that are not crystalline or are treated as defects, taken over the whole sample. Its boundary surface $\Sigma$ is made of the
Voronoi faces, edges, and vertices shared by atoms both in $D$ and outside it: mixed pairs (faces), triples (edges),
and quadruples (vertices) of atoms. In a generic configuration (three cells at every Voronoi edge, four at every
vertex), as thermal snapshots are, each mixed edge lies on exactly two mixed faces, so $\Sigma$ is a closed surface.
Each connected piece of $\Sigma$ has Euler characteristic $\chi=V-E+F$, with $V$, $E$, and $F$ the numbers of vertices, edges, and
faces of that piece, and genus $h=1-\chi/2$ ($\chi$ is always even).

\emph{Curves.} On each piece, a standard construction gives $2h$ curves, two per handle, from which every
closed curve on the surface can be composed~\cite{eppstein2003} (Supplementary Section~\ref{sec:si-alg}). Curves that
can be shrunk to a point through the crystal outside the cluster return zero. A curve, a
cyclic sequence of mixed pairs $\{d,x\}$ with $d\in D$, becomes the walk through the shell atoms $x$, with
consecutive repeats removed. Consecutive shell atoms share a face, so every step has a frame relation (for
close-packed crystals, see below).

\emph{Burgers vectors.} Each region of crystalline cells has its own frame, so curves are combined only within one region. A
curve's winding around the periodic box is the sum of its minimum-image steps in fractional coordinates of the
(possibly triclinic) box. Curves that wind around the box enter only through integer combinations with zero total
winding, found by exact integer elimination. These combinations give the cluster's Burgers vectors, which are lattice
vectors (Supplementary Section~\ref{sec:si-alg}). We report the vectors of the cluster's shortest closed curves, up to sign, and generators of the lattice they span.
The sign is left open, since a closed curve, like a dislocation line, has no preferred direction. In the MD tests with straight
lines it was read from each curve's winding around the inserted lines.

\emph{Burgers vector of each core atom.} A cluster that contains a dislocation network has one vector per branch,
so each atom takes the vector of the branch it lies on. If the cluster has a single vector, every atom takes it. Otherwise, each shell
atom $x$ takes the net translation, up to sign, of the shortest closed walk through $x$ within the shell whose net
translation is not zero and which does not wind around the box. Such a walk runs once around the branch on which $x$
lies. Where walks of the same least length give different vectors, $x$ takes all of them. Each atom of $D$ takes the
vectors of the shell atoms nearest to it, with distances counted in steps between face-sharing atoms through the
cluster, and again keeps all vectors that tie. In our samples, ties occur only near junctions. In drawings, an atom with several vectors takes one of them. The assignment does not depend on the choice of starting atom or the order of search. It can be computed exactly
because the frames of the shell atoms can be fixed once, along a spanning tree, so that the net translation of a walk
is a sum of integer vectors, one per step (Supplementary Section~\ref{sec:si-alg}).

\emph{Circuits for figures.} The Burgers circuits drawn in Figs.~\ref{fig:concept} and \ref{fig:gallery}
are open walks through crystalline atoms whose steps, taken as ideal interatomic vectors, sum to
zero; one step equal to $\bvec$ closes each (Supplementary Section~\ref{sec:si-alg}). They illustrate the result; every reported vector
comes from the boundary surface.

\subsection*{Close-packed crystals}
For fcc and hcp crystals, a table of cell topologies tells which faces of a cell join its full neighbors. A small
perturbation of the ideal crystal changes a cell's topology only at its degenerate vertices (the vertices shared by more than four cells, each of which can split in multiple ways; Supplementary Section~\ref{sec:si-alg}), so we enumerate the splittings of these vertices around one
atom, and all compatible combinations up to the site symmetry, as in VoroTypes~\cite{votypes2026b}. For each
resulting topology the table records the sets of its faces that join the twelve full neighbors of the ideal crystal,
over all isomorphisms with the enumerated cells. The fcc table has 2{,}815 topologies and the hcp table 9{,}556. Supplementary
Section~\ref{sec:si-degenerate} shows the same construction in two dimensions.
\emph{Agreement.} The table gives each cell a list of possible alignments. Two neighboring cells must agree on their
shared face: it joins full neighbors for both cells or for neither. If an alignment of a cell disagrees with every
remaining alignment of one of its neighbors, we discard it. We repeat this until nothing more can be discarded (arc
consistency~\cite{mackworth1977}). A cell whose topology is not in the table, or that has no alignment left, is
treated as a defect. Its neighbors then ignore it, and agreement starts again without it. \emph{Checks.} A cell left with a
single alignment is treated as a defect unless its twelve full neighbors form their ideal arrangement. Each pair of the twelve that should be full neighbors of each other must be confirmed as
such by at least one of the two that itself has a single alignment. Matching the twelve to this arrangement gives each its ideal vector, and hence the
cell's labeling. The cell is also treated as a defect unless its frame relation with every crystalline cell it shares a face with is
a crystal symmetry. Atoms treated as defects join the defect clusters. Consecutive atoms of a boundary curve that are only partial neighbors are related
through crystalline atoms that neighbor both (Supplementary Section~\ref{sec:si-alg}).

\subsection*{Other degenerate crystals}
For $\omega$-Ti, unlike fcc and hcp, the perturbed cells serve directly as reference cells. For each site we list
with VoroTypes every way its degenerate vertices can split, up to the site symmetry, and keep the resulting cells,
each with the ideal neighbor vector on every face. A topology can occur with different assignments of neighbor vectors to its faces, not related by a symmetry
of the site, and each is kept as a separate labeling. Of the 388 topologies (8 at the 1a site and 382 at the 2d site,
two of them at both), 384 have a single labeling, three have two, and one has nine, for 399 labelings in all. Each cell
of a sample is then labeled by matching it to one of these, as in a
stable crystal. One difference remains. Near a partial neighbor, the labeling does not determine the ideal positions
of a perturbed cell's vertices, so frame relations are obtained not from the vertices of the shared face but from the
ideal positions both labelings give to the two atoms of the face and their common neighbors.

\subsection*{Crystal structures}
For each Voronoi vertex of the ideal crystal in three dimensions, let $r_4$ and $r_5$ be its distances to the fourth and fifth nearest
atoms. The Voronoi stability gap $g$ of the crystal is the minimum of $(r_5-r_4)/r_4$ over all vertices. If $g>0$, every vertex is shared by exactly four cells (the crystal is topologically stable), so every cell has a single topology, stable under
small displacements. We computed $g$ from the stored coordinates
(Materials Project database version 2026.09.28, all structures ordered). VoroTypes~\cite{votypes2026a,votypes2026b}, an independent implementation, found a single cell topology
per Wyckoff site and the same gap, to three decimals, in all 9{,}610 structures of the test. Topology families are
sets of distinct reference cell topologies; geometries were grouped with pymatgen's structure matcher~\cite{ong2013pymatgen}, ignoring
species.

Each structure was first snapped onto exact positions of its Materials Project space group, at the smallest
tolerance ($10^{-5}$ to $0.2$~\AA) at which spglib~\cite{spglib} finds that group, as in VoroTypes~\cite{votypes2026a}. For 9
structures, no tolerance in this range recovers the declared group, and their stored coordinates were used unchanged. Snapping changes the primitive cell of 34
structures, whose files list a supercell of the crystal (for example, a seven-atom triclinic cell of a one-atom
rhombohedral Nb). The analysis then sees the crystal as the database classifies it, and its results can depend on
that classification, as they depend on the choice between an ordered reference and a species-blind one (Results).

Each reference was built from the snapped structure alone; its lattice was LLL-reduced~\cite{lll1982} (reduced to a
basis of short, nearly orthogonal vectors), with the shortest lattice vector $\mathbf s$ as the cell vector
$\mathbf c$, and written in restricted-triclinic form (the LAMMPS convention). The supercells are fully periodic, at least $28d$ wide in
cross-section ($56d$ or $112d$ where needed; Results) and at least $6d$ long, with $d$ the nearest-neighbor distance. Whether a curve wraps the box is read from its steps, each taken to the nearest
periodic image, so the sample must be longer along the line than twice the longest neighbor distance. A length of
$6d$ satisfies this in every one of the 9{,}610 structures. Two dipoles were inserted: a screw dipole with
$\bvec=\mathbf s$, and a non-screw dipole with $\bvec$ the shortest lattice vector not parallel to $\mathbf s$. The two
lines of the non-screw dipole are separated along the edge component of $\bvec$, so that the cut between them contains
$\bvec$ (a cut across it would add or remove a sheet of atoms). A third dipole had its lines along that vector and
$\bvec=\mathbf s$. The displacement is the textbook elastic field of a straight dislocation in an isotropic
medium~\cite{hirthlothe}, summed for the dipole over its periodic images in the cross-section. It uses a Poisson ratio
of 0.3 for every structure, since only the Burgers vector it inserts matters for the test. The residual uniform gradient of the finite sum is removed, so that the sample
is periodic to within 0.002~\AA. Gaussian noise of standard deviation $0.02\,d$ per Cartesian component was then added,
to mimic thermal displacements.

Each Burgers vector is computed exactly in the frame of the crystal (Supplementary Section~\ref{sec:si-alg}). To compare it with the
inserted vector, it is rotated into the laboratory frame, using the crystal's orientation, which the software fits to the
neighbor vectors, and replaced by the nearest lattice vector. This is correct as long as the rotated vector is off by less than
half the shortest lattice vector, whose length is at least $d$. The largest error in the whole test was $0.22\,d$, well within
this margin. A dipole was counted as correct if some cluster returns a
vector and every vector found is $\pm\bvec$, as wrong if any other vector
appears, and as undecided if no cluster returns a vector. Undecided dipoles, and dipoles whose two lines merged into
one cluster, were analyzed again in cross-sections twice and four times as wide.

\subsection*{Molecular dynamics}
Simulations used LAMMPS~\cite{lammps} with embedded-atom potentials for Fe~\cite{mendelev2003},
Cu~\cite{mishin2001}, Mg~\cite{sun2006}, NiAl~\cite{mishin2002}, and MgCu$_2$ (from the database of Zhou, Johnson, and Wadley~\cite{zhou2004eam}), the angular-dependent potential of Apostol and Mishin for
Al--Cu~\cite{apostol2011}, and the spline-based modified embedded-atom potential of Hennig et al.\ for
Ti~\cite{hennig2008}. \emph{Validation samples.} Screw dislocations form a checkerboard, and edge and mixed ones the standard
$45^\circ$ quadrupole, with partners on different glide planes. The displacement is the same elastic field as in the Materials
Project test, summed over periodic images and superposed for the four lines. Where a cut has a climb
component, the atoms it doubles are removed by an exact rule decided on the ideal crystal: atom $i+\bvec$ is
removed when the segment from $i$ to $i+\bvec$ crosses the cut. This removes exactly the slab volume times the
density of each species. The prismatic loop was built by removing a disc of three (111) layers and closing the gap
with the solid-angle displacement field~\cite{hirthlothe}, which jumps by exactly $\bvec$ across the disc. Cross-sections are 59--285~\AA. Samples
were minimized with the box relaxed, run for 15~ps at constant temperature and zero pressure (NPT), and then, with the
box fixed at its average, at constant temperature (NVT, Nos\'e--Hoover thermostat). After 5~ps, raw snapshots were taken every 1~ps (1.5~ps for $\theta$, 3~ps for Mg). Wherever the
lines of a quadrupole annihilated or merged during the run, the cell was doubled. In
$\omega$-Ti the lines annihilated after about 18~ps of this protocol, so we analyzed a run at constant pressure started
from the minimized sample, every 1~ps from 1 to 19~ps (Supplementary Section~\ref{sec:si-degenerate}). The close-packed analysis currently requires orthogonal boxes, so Cu and Mg were relaxed
with the box kept orthogonal. DXA analyses used OVITO 3.17~\cite{ovito} with default parameters unless stated otherwise. The Fe potential's melting temperature, $T_\mathrm m=1770$~K, was found by solid--liquid coexistence
in a box of $6.4\times10^4$ atoms (Supplementary Section~\ref{sec:si-val}).

\emph{Plastic flow (Fig.~\ref{fig:showcase}).} The samples of $\theta$-Al$_2$Cu
(Apostol--Mishin potential) and C15 MgCu$_2$ (embedded-atom potential~\cite{zhou2004eam}) were seeded with circular glide
loops. For C15, they lie on the four $\{111\}$ planes and on $\{001\}$, all with
$\tfrac12\langle110\rangle$ vectors. For $\theta$, they cover the five most favorable systems identified by Andre et
al.~\cite{andre2021theta}: $\{211\}\tfrac12\langle111\rangle$ (two loops), $\{011\}\tfrac12\langle111\rangle$,
$\{011\}[100]$, $\{110\}[001]$, and $\{310\}[001]$. Each loop was inserted by the solid-angle
displacement field. After energy minimization and 10~ps at 300~K (Nos\'e--Hoover thermostat~\cite{nose1984, hoover1985}), the box was
deformed at a constant strain rate of $5\times10^{8}$~s$^{-1}$ for 200~ps (normal components applied to the box
lengths, shear components as tilts, with atoms remapped), a total strain of 10\%. The strain tensor, with normal and
shear components, and the sign of each loop were chosen so that every seeded loop expands. Snapshots were written every 0.2~ps ($\theta$) and 1~ps
(C15) and analyzed directly, without minimization or averaging, in their triclinic boxes.

\emph{Junction (Fig.~\ref{fig:junction}).} Two circular loops of radius 60~\AA, on $(2\bar1\bar1)$ with
$\bvec=\tfrac12[111]$ and on $(211)$ with $\bvec=\tfrac12[1\bar1\bar1]$, were inserted in a fully periodic
$\theta$-Al$_2$Cu box of 2{,}274{,}480 atoms by the solid-angle displacement field, tapered to zero far from the loops
so that the field is periodic. The planes meet along $[01\bar1]$, the line of a $[100]$ junction. In one sample both loops
pass through the same point of this line, at the center of the box; in the others each loop stops 8 or 11~\AA\ short of it,
within its own plane. After energy minimization, the samples were run at 300~K (Nos\'e--Hoover) with no applied strain
for 200~ps, the 11~\AA\ sample for 50~ps. The sample with crossing lines was continued at 300~K to 359~ps. A copy of the
same initial sample, minimized again, was held at 150~K for 50~ps and then heated to 300, 450, and 600~K, with
25~ps ramps and 50~ps holds. All runs kept the box fixed. Snapshots were written every 1~ps, except in the 8 and 11~\AA\ samples (every 0.2~ps for the
first 50~ps, and every 0.5~ps afterward) and in the copy heated in steps (every 0.5~ps, and every 0.25~ps during heating).

\subsection*{Implementation}
The analysis, for crystals with stable cells and for fcc and hcp, is implemented in C++ on top of voro++, with OpenMP
through the multithreaded voro++ containers~\cite{lu2023voromt}. A Python package provides the defect clusters, their
surfaces and loops, approximate line positions for display, and a command-line tool. For degenerate crystals such as $\omega$-Ti,
VoroTypes~\cite{votypes2026b} enumerates the perturbed cells.

\subsection*{Run time and memory}
The stages applied to every atom take time proportional to the number of atoms~$N$, up to sorting: each cell has a
bounded number of faces and labelings, and arc consistency revisits a cell only when a neighbor loses a labeling. The
cluster analysis involves only the atoms in and around defect clusters. We timed the analysis, from the positions in
memory to the Burgers vectors of all clusters, on molecular-dynamics snapshots tiled periodically to 1.2--79 million
atoms, on four types of node, with DXA (OVITO 3.17, default parameters, with no template for $\theta$-Al$_2$Cu or C15)
on the same samples, node, and threads (Table~\ref{tab:runtime}; Supplementary Section~\ref{sec:si-runtime}). With 16 threads the analysis takes
6--10~s per million atoms for bcc Fe and $\theta$-Al$_2$Cu, 8--11~s for fcc Cu, and 13--18~s for hcp Mg. For bcc Fe
from 1.2 to 79 million atoms, the time grows about linearly in $N$ (fitted exponents 0.88--1.05 on the four nodes), and peak memory is
1.4--1.5~GB per million atoms, against 1.1--1.3~GB for DXA. On one thread DXA is 1.2--2.3 times faster, but our analysis gains
more from threads (5--7 times from 1 to 16 threads on a Gold 6348 node, against 1.5 times for DXA),
and with 16 threads it takes 0.4--0.8 times the time of DXA on bcc Fe and fcc Cu, and 0.9--1.1 times on hcp Mg. Since DXA also traces and smooths lines, these comparisons show
only that the analysis is practical at the sizes where DXA is used. It is slowest when one defect cluster holds much of
the sample: in the C15 network (12\% of the atoms not crystalline), the cluster analysis, in Python, takes
about half the time, and the time per atom about doubles from 1.2 to 9.7 million atoms on Gold nodes.

\begin{table}[!htbp]
\centering\footnotesize
\begin{tabular}{lrrrrr}
\toprule
 & atoms & \multicolumn{2}{c}{time (s)} & \multicolumn{2}{c}{memory (GB)}\\
sample & ($10^6$) & ours & DXA & ours & DXA\\
\midrule
bcc Fe, prismatic loops & 1.24 & 7.8 & 15.1 & 1.9 & 1.6\\
 & 9.90 & 66.5 & 113.4 & 13.7 & 11.1\\
 & 79.17 & 591.8 & 915.3 & 108.7 & 84.9\\
bcc Fe, screw quadrupole & 1.55 & 9.7 & 18.6 & 2.3 & 2.0\\
bcc Fe, edge quadrupole & 2.12 & 14.1 & 25.7 & 3.0 & 2.6\\
$\theta$-Al$_2$Cu, junction & 2.27 & 14.7 & -- & 3.4 & --\\
 & 18.20 & 129.0 & -- & 24.7 & --\\
C15 MgCu$_2$, network & 1.22 & 15.0 & -- & 2.0 & --\\
 & 9.73 & 264.1 & -- & 19.7 & --\\
fcc Cu, dissociated edges & 1.94 & 17.9 & 27.1 & 3.3 & 2.5\\
 & 15.53 & 145.6 & 198.8 & 22.5 & 17.7\\
hcp Mg, $[0001]$ screws & 1.32 & 18.1 & 18.0 & 2.8 & 1.7\\
 & 10.58 & 150.6 & 135.1 & 16.6 & 12.1\\
\bottomrule
\end{tabular}
\caption{Run time and peak memory for one snapshot, with 16 threads on a Xeon Gold 6348 node.}
\label{tab:runtime}
\end{table}

\subsection*{Use of AI tools}
Anthropic's Claude Code, a coding assistant based on a large language model, was used extensively throughout this
work, in algorithm development, code implementation, computational verification, and manuscript drafting. The author
directed the work, checked the results, and takes full responsibility for the content.

\subsection*{Code availability}
The analysis is part of VoroTop 1.2 (BSD 3-clause license, \url{https://github.com/VoroTop/VoroTop}). The scripts used
for this paper are archived with the data.

\subsection*{Data availability}
The 9{,}610 tested structures (Materials Project identifiers, gaps, families, and results), all molecular-dynamics inputs
and initial configurations, the analysis outputs of every snapshot, and the snapshots shown in the figures are
on Zenodo (\url{https://doi.org/10.5281/zenodo.24302060}).

\section*{Acknowledgments}
This paper is dedicated to the memory of Robert V. Kohn (1953--2026) and Frank H. Lutz (1968--2023).

\medskip\noindent The author acknowledges support from the Israel Science Foundation under grant 202/24.

\section*{Author contributions}
E.A.L. conceived the study, carried out the research, and wrote the manuscript.

\section*{Competing interests}
The author declares no competing interests.

\bibliographystyle{unsrtnat}
\bibliography{refs}

\begin{thebibliography}{58}
\providecommand{\natexlab}[1]{#1}
\providecommand{\url}[1]{\texttt{#1}}
\expandafter\ifx\csname urlstyle\endcsname\relax
  \providecommand{\doi}[1]{doi: #1}\else
  \providecommand{\doi}{doi: \begingroup \urlstyle{rm}\Url}\fi

\bibitem[Zepeda-Ruiz et~al.(2021)Zepeda-Ruiz, Stukowski, Oppelstrup, Bertin,
  Barton, Freitas, and Bulatov]{zepeda2021}
L.~A. Zepeda-Ruiz, A.~Stukowski, T.~Oppelstrup, N.~Bertin, N.~R. Barton,
  R.~Freitas, and V.~V. Bulatov.
\newblock Atomistic insights into metal hardening.
\newblock \emph{Nat. Mater.}, 20:\penalty0 315--320, 2021.
\newblock \doi{10.1038/s41563-020-00815-1}.

\bibitem[Stukowski and Albe(2010{\natexlab{a}})]{stukowski2010dxa}
A.~Stukowski and K.~Albe.
\newblock Dislocation detection algorithm for atomistic simulations.
\newblock \emph{Modelling Simul. Mater. Sci. Eng.}, 18:\penalty0 025016,
  2010{\natexlab{a}}.
\newblock \doi{10.1088/0965-0393/18/2/025016}.

\bibitem[Stukowski and Albe(2010{\natexlab{b}})]{stukowski2010defects}
A.~Stukowski and K.~Albe.
\newblock Extracting dislocations and non-dislocation crystal defects from
  atomistic simulation data.
\newblock \emph{Modelling Simul. Mater. Sci. Eng.}, 18:\penalty0 085001,
  2010{\natexlab{b}}.
\newblock \doi{10.1088/0965-0393/18/8/085001}.

\bibitem[Stukowski et~al.(2012)Stukowski, Bulatov, and
  Arsenlis]{stukowski2012dxa}
A.~Stukowski, V.~V. Bulatov, and A.~Arsenlis.
\newblock Automated identification and indexing of dislocations in crystal
  interfaces.
\newblock \emph{Modelling Simul. Mater. Sci. Eng.}, 20:\penalty0 085007, 2012.
\newblock \doi{10.1088/0965-0393/20/8/085007}.

\bibitem[Stukowski(2014)]{stukowski2014jmps}
A.~Stukowski.
\newblock A triangulation-based method to identify dislocations in atomistic
  models.
\newblock \emph{J. Mech. Phys. Solids}, 70:\penalty0 314--319, 2014.
\newblock \doi{10.1016/j.jmps.2014.06.009}.

\bibitem[Honeycutt and Andersen(1987)]{honeycutt1987}
J.~D. Honeycutt and H.~C. Andersen.
\newblock Molecular dynamics study of melting and freezing of small
  {Lennard-Jones} clusters.
\newblock \emph{J. Phys. Chem.}, 91:\penalty0 4950--4963, 1987.
\newblock \doi{10.1021/j100303a014}.

\bibitem[Faken and J{\'o}nsson(1994)]{faken1994}
D.~Faken and H.~J{\'o}nsson.
\newblock Systematic analysis of local atomic structure combined with {3D}
  computer graphics.
\newblock \emph{Comput. Mater. Sci.}, 2:\penalty0 279--286, 1994.
\newblock \doi{10.1016/0927-0256(94)90109-0}.

\bibitem[Stukowski(2012)]{stukowski2012cna}
A.~Stukowski.
\newblock Structure identification methods for atomistic simulations of
  crystalline materials.
\newblock \emph{Modelling Simul. Mater. Sci. Eng.}, 20:\penalty0 045021, 2012.
\newblock \doi{10.1088/0965-0393/20/4/045021}.

\bibitem[Hartley and Mishin(2005)]{hartley2005}
C.~S. Hartley and Y.~Mishin.
\newblock Characterization and visualization of the lattice misfit associated
  with dislocation cores.
\newblock \emph{Acta Mater.}, 53:\penalty0 1313--1321, 2005.
\newblock \doi{10.1016/j.actamat.2004.11.027}.

\bibitem[Dai and Zhang(2015)]{dai2015}
F.~Z. Dai and W.~Z. Zhang.
\newblock An automatic and simple method for specifying dislocation features in
  atomistic simulations.
\newblock \emph{Comput. Phys. Commun.}, 188:\penalty0 103--109, 2015.
\newblock \doi{10.1016/j.cpc.2014.11.014}.

\bibitem[Zimmerman et~al.(2001)Zimmerman, Kelchner, Klein, Hamilton, and
  Foiles]{zimmerman2001}
J.~A. Zimmerman, C.~L. Kelchner, P.~A. Klein, J.~C. Hamilton, and S.~M. Foiles.
\newblock Surface step effects on nanoindentation.
\newblock \emph{Phys. Rev. Lett.}, 87:\penalty0 165507, 2001.
\newblock \doi{10.1103/PhysRevLett.87.165507}.

\bibitem[Hua and Hartmaier(2010)]{hua2010}
J.~Hua and A.~Hartmaier.
\newblock Determining {Burgers} vectors and geometrically necessary dislocation
  densities from atomistic data.
\newblock \emph{Modelling Simul. Mater. Sci. Eng.}, 18:\penalty0 045007, 2010.
\newblock \doi{10.1088/0965-0393/18/4/045007}.

\bibitem[Bhardwaj(2026)]{bhardwaj2026savi}
U.~Bhardwaj.
\newblock {Savi-Bhransha}: Graph-theoretic dislocation-loop characterization in
  crystals.
\newblock \emph{arXiv:2607.29328}, 2026.
\newblock \doi{10.48550/arXiv.2607.29328}.

\bibitem[ovi(2026)]{ovitodxa}
Dislocation analysis ({DXA}).
\newblock OVITO 3 user manual,
  \url{https://www.ovito.org/docs/current/reference/pipelines/modifiers/dislocation_analysis.html},
  2026.

\bibitem[Xie et~al.(2021)Xie, Chauraud, Bitzek, Korte-Kerzel, and
  Gu{\'e}nol{\'e}]{laca2021}
Z.~Xie, D.~Chauraud, E.~Bitzek, S.~Korte-Kerzel, and J.~Gu{\'e}nol{\'e}.
\newblock Laves phase crystal analysis ({LaCA}): Atomistic identification of
  lattice defects in {C14} and {C15} topologically close-packed phases.
\newblock \emph{J. Mater. Res.}, 36:\penalty0 2010--2024, 2021.
\newblock \doi{10.1557/s43578-021-00237-y}.

\bibitem[Okafor et~al.(2025)Okafor, Sayyadi-Shahraki, Bruns, Fr{\"o}mling,
  Hirel, Carrez, Durst, and Fang]{okafor2025srtio3}
C.~Okafor, A.~Sayyadi-Shahraki, S.~Bruns, T.~Fr{\"o}mling, P.~Hirel, P.~Carrez,
  K.~Durst, and X.~Fang.
\newblock Coupled electromigration-nanoindentation study on dislocation
  nucleation in {SrTiO$_3$}.
\newblock \emph{J. Am. Ceram. Soc.}, 108\penalty0 (10):\penalty0 e70015, 2025.
\newblock \doi{10.1111/jace.70015}.

\bibitem[Xu et~al.(2024)Xu, Zhang, Bitzek, Schmauder, and
  Grabowski]{xu2024ni3al}
X.~Xu, X.~Zhang, E.~Bitzek, S.~Schmauder, and B.~Grabowski.
\newblock Origin of the yield stress anomaly in {L1$_2$} intermetallics
  unveiled with physically informed machine-learning potentials.
\newblock \emph{Acta Mater.}, 281:\penalty0 120423, 2024.
\newblock \doi{10.1016/j.actamat.2024.120423}.

\bibitem[Hirel et~al.(2014)Hirel, Kraych, Carrez, and Cordier]{hirel2014}
P.~Hirel, A.~Kraych, P.~Carrez, and P.~Cordier.
\newblock Atomic core structure and mobility of [100](010) and [010](100)
  dislocations in {MgSiO$_3$} perovskite.
\newblock \emph{Acta Mater.}, 79:\penalty0 117--125, 2014.
\newblock \doi{10.1016/j.actamat.2014.07.001}.

\bibitem[Freund et~al.(2026)Freund, Spille, Sun, Lipi{\'n}ska-Chwa{\l}ek,
  Mayer, Xie, and Korte-Kerzel]{freund2026}
M.~Freund, J.~Spille, P.-L. Sun, M.~Lipi{\'n}ska-Chwa{\l}ek, J.~Mayer, Z.~Xie,
  and S.~Korte-Kerzel.
\newblock Redefining plasticity in a cubic intermetallic: Dislocation dynamics
  in the {CaAl$_2$} {Laves} phase.
\newblock \emph{Acta Mater.}, 303:\penalty0 121604, 2026.
\newblock \doi{10.1016/j.actamat.2025.121604}.

\bibitem[Zhang et~al.(2024)Zhang, Li, Suzuki, Nakamura, and
  Ogata]{zhang2024ceramics}
S.~Zhang, Y.~Li, S.~Suzuki, A.~Nakamura, and S.~Ogata.
\newblock Neural network potential for dislocation plasticity in ceramics.
\newblock \emph{npj Comput. Mater.}, 10:\penalty0 266, 2024.
\newblock \doi{10.1038/s41524-024-01456-7}.

\bibitem[Zhou et~al.(2017)Zhou, Wang, Misra, Huang, Wang, and
  Xu]{zhou2017al2cu}
Q.~Zhou, J.~Wang, A.~Misra, P.~Huang, F.~Wang, and K.~Xu.
\newblock Dislocations interaction induced structural instability in
  intermetallic {Al$_2$Cu}.
\newblock \emph{npj Comput. Mater.}, 3:\penalty0 24, 2017.
\newblock \doi{10.1038/s41524-017-0030-2}.

\bibitem[Landweber et~al.(2016)Landweber, Lazar, and Patel]{landweber2016}
P.~S. Landweber, E.~A. Lazar, and N.~Patel.
\newblock On fiber diameters of continuous maps.
\newblock \emph{Am. Math. Mon.}, 123\penalty0 (4):\penalty0 392--397, 2016.
\newblock \doi{10.4169/amer.math.monthly.123.4.392}.

\bibitem[Lazar et~al.(2022)Lazar, Lu, and Rycroft]{lazar2022ajp}
E.~A. Lazar, J.~Lu, and C.~H. Rycroft.
\newblock Voronoi cell analysis: The shapes of particle systems.
\newblock \emph{Am. J. Phys.}, 90\penalty0 (6):\penalty0 469--480, 2022.
\newblock \doi{10.1119/5.0087591}.

\bibitem[Lazar et~al.(2015)Lazar, Han, and Srolovitz]{lazar2015pnas}
E.~A. Lazar, J.~Han, and D.~J. Srolovitz.
\newblock Topological framework for local structure analysis in condensed
  matter.
\newblock \emph{Proc. Natl. Acad. Sci. USA}, 112:\penalty0 E5769--E5776, 2015.
\newblock \doi{10.1073/pnas.1505788112}.

\bibitem[Lazar(2018)]{vorotop2022}
E.~A. Lazar.
\newblock {VoroTop}: {Voronoi} cell topology visualization and analysis
  toolkit.
\newblock \emph{Modelling Simul. Mater. Sci. Eng.}, 26:\penalty0 015011, 2018.
\newblock \doi{10.1088/1361-651X/aa9a01}.

\bibitem[Lazar and Rycroft(2026{\natexlab{a}})]{votypes2026a}
E.~A. Lazar and C.~H. Rycroft.
\newblock Structural identification in general crystalline systems via
  {Voronoi} cell topology.
\newblock \emph{arXiv:2610.20498}, 2026{\natexlab{a}}.
\newblock \doi{10.48550/arXiv.2610.20498}.

\bibitem[Lazar and Rycroft(2026{\natexlab{b}})]{votypes2026b}
E.~A. Lazar and C.~H. Rycroft.
\newblock {VoroTypes}: automated enumeration of {Voronoi} cell topologies for
  structural identification of complex crystals in atomistic simulations.
\newblock \emph{arXiv:2610.21911}, 2026{\natexlab{b}}.
\newblock \doi{10.48550/arXiv.2610.21911}.

\bibitem[Hirth and Lothe(1982)]{hirthlothe}
J.~P. Hirth and J.~Lothe.
\newblock \emph{Theory of Dislocations}.
\newblock Wiley, 2nd edition, 1982.

\bibitem[Lazar et~al.(2024)Lazar, Lu, Rycroft, and Schwarcz]{lazar2024twod}
E.~A. Lazar, J.~Lu, C.~H. Rycroft, and D.~Schwarcz.
\newblock Characterizing structural features of two-dimensional particle
  systems through {Voronoi} topology.
\newblock \emph{Modelling Simul. Mater. Sci. Eng.}, 32:\penalty0 085022, 2024.
\newblock \doi{10.1088/1361-651X/ad8ad9}.

\bibitem[Nelson and Halperin(1979)]{nelson1979}
D.~R. Nelson and B.~I. Halperin.
\newblock Dislocation-mediated melting in two dimensions.
\newblock \emph{Phys. Rev. B}, 19:\penalty0 2457--2484, 1979.
\newblock \doi{10.1103/PhysRevB.19.2457}.

\bibitem[Eppstein(2003)]{eppstein2003}
D.~Eppstein.
\newblock Dynamic generators of topologically embedded graphs.
\newblock In \emph{Proceedings of the 14th Annual ACM-SIAM Symposium on
  Discrete Algorithms (SODA)}, pages 599--608, 2003.

\bibitem[Jain et~al.(2013)Jain, Ong, Hautier, Chen, Richards, Dacek, Cholia,
  Gunter, Skinner, Ceder, and Persson]{jain2013mp}
A.~Jain, S.~P. Ong, G.~Hautier, W.~Chen, W.~D. Richards, S.~Dacek, S.~Cholia,
  D.~Gunter, D.~Skinner, G.~Ceder, and K.~A. Persson.
\newblock Commentary: The {Materials Project}: A materials genome approach to
  accelerating materials innovation.
\newblock \emph{APL Mater.}, 1:\penalty0 011002, 2013.
\newblock \doi{10.1063/1.4812323}.

\bibitem[Mendelev et~al.(2003)Mendelev, Han, Srolovitz, Ackland, Sun, and
  Asta]{mendelev2003}
M.~I. Mendelev, S.~Han, D.~J. Srolovitz, G.~J. Ackland, D.~Y. Sun, and M.~Asta.
\newblock Development of new interatomic potentials appropriate for crystalline
  and liquid iron.
\newblock \emph{Philos. Mag.}, 83:\penalty0 3977--3994, 2003.
\newblock \doi{10.1080/14786430310001613264}.

\bibitem[Stukowski(2010)]{ovito}
A.~Stukowski.
\newblock Visualization and analysis of atomistic simulation data with
  {OVITO}--the {Open Visualization Tool}.
\newblock \emph{Modelling Simul. Mater. Sci. Eng.}, 18:\penalty0 015012, 2010.
\newblock \doi{10.1088/0965-0393/18/1/015012}.

\bibitem[Xie et~al.(2023)Xie, Chauraud, Atila, Bitzek, Korte-Kerzel, and
  Gu{\'e}nol{\'e}]{xie2023synchroshockley}
Z.~Xie, D.~Chauraud, A.~Atila, E.~Bitzek, S.~Korte-Kerzel, and
  J.~Gu{\'e}nol{\'e}.
\newblock Unveiling the mechanisms of motion of synchro-{Shockley} dislocations
  in {Laves} phases.
\newblock \emph{Phys. Rev. Mater.}, 7:\penalty0 053605, 2023.
\newblock \doi{10.1103/PhysRevMaterials.7.053605}.

\bibitem[Larsen et~al.(2016)Larsen, Schmidt, and Schi{\o}tz]{larsen2016ptm}
P.~M. Larsen, S.~Schmidt, and J.~Schi{\o}tz.
\newblock Robust structural identification via polyhedral template matching.
\newblock \emph{Modelling Simul. Mater. Sci. Eng.}, 24:\penalty0 055007, 2016.
\newblock \doi{10.1088/0965-0393/24/5/055007}.

\bibitem[Bilby et~al.(1955)Bilby, Bullough, and Smith]{bilby1955}
B.~A. Bilby, R.~Bullough, and E.~Smith.
\newblock Continuous distributions of dislocations: a new application of the
  methods of non-{Riemannian} geometry.
\newblock \emph{Proc. R. Soc. Lond. A}, 231:\penalty0 263--273, 1955.
\newblock \doi{10.1098/rspa.1955.0171}.

\bibitem[Kl{\'e}man and Friedel(2008)]{kleman2008}
M.~Kl{\'e}man and J.~Friedel.
\newblock Disclinations, dislocations, and continuous defects: A reappraisal.
\newblock \emph{Rev. Mod. Phys.}, 80:\penalty0 61--115, 2008.
\newblock \doi{10.1103/RevModPhys.80.61}.

\bibitem[Rycroft(2009)]{voro}
C.~H. Rycroft.
\newblock {VORO++}: A three-dimensional {Voronoi} cell library in {C++}.
\newblock \emph{Chaos}, 19:\penalty0 041111, 2009.
\newblock \doi{10.1063/1.3215722}.

\bibitem[Lu et~al.(2023)Lu, Lazar, and Rycroft]{lu2023voromt}
J.~Lu, E.~A. Lazar, and C.~H. Rycroft.
\newblock An extension to {VORO++} for multithreaded computation of {Voronoi}
  cells.
\newblock \emph{Comput. Phys. Commun.}, 291:\penalty0 108832, 2023.
\newblock \doi{10.1016/j.cpc.2023.108832}.

\bibitem[Togo et~al.(2024)Togo, Shinohara, and Tanaka]{spglib}
A.~Togo, K.~Shinohara, and I.~Tanaka.
\newblock Spglib: a software library for crystal symmetry search.
\newblock \emph{Sci. Technol. Adv. Mater. Methods}, 4:\penalty0 2384822, 2024.
\newblock \doi{10.1080/27660400.2024.2384822}.

\bibitem[Weinberg(1966)]{weinberg1966}
L.~Weinberg.
\newblock A simple and efficient algorithm for determining isomorphism of
  planar triply connected graphs.
\newblock \emph{IEEE Trans. Circuit Theory}, 13:\penalty0 142--148, 1966.
\newblock \doi{10.1109/TCT.1966.1082573}.

\bibitem[Lazar et~al.(2012)Lazar, Mason, MacPherson, and
  Srolovitz]{lazar2012prl}
E.~A. Lazar, J.~K. Mason, R.~D. MacPherson, and D.~J. Srolovitz.
\newblock Complete topology of cells, grains, and bubbles in three-dimensional
  microstructures.
\newblock \emph{Phys. Rev. Lett.}, 109:\penalty0 095505, 2012.
\newblock \doi{10.1103/PhysRevLett.109.095505}.

\bibitem[Mackworth(1977)]{mackworth1977}
A.~K. Mackworth.
\newblock Consistency in networks of relations.
\newblock \emph{Artif. Intell.}, 8:\penalty0 99--118, 1977.
\newblock \doi{10.1016/0004-3702(77)90007-8}.

\bibitem[Ong et~al.(2013)Ong, Richards, Jain, Hautier, Kocher, Cholia, Gunter,
  Chevrier, Persson, and Ceder]{ong2013pymatgen}
S.~P. Ong, W.~D. Richards, A.~Jain, G.~Hautier, M.~Kocher, S.~Cholia,
  D.~Gunter, V.~L. Chevrier, K.~A. Persson, and G.~Ceder.
\newblock {Python Materials Genomics} (pymatgen): A robust, open-source
  {Python} library for materials analysis.
\newblock \emph{Comput. Mater. Sci.}, 68:\penalty0 314--319, 2013.
\newblock \doi{10.1016/j.commatsci.2012.10.028}.

\bibitem[Lenstra et~al.(1982)Lenstra, Lenstra, and Lov{\'a}sz]{lll1982}
A.~K. Lenstra, H.~W. Lenstra, and L.~Lov{\'a}sz.
\newblock Factoring polynomials with rational coefficients.
\newblock \emph{Math. Ann.}, 261:\penalty0 515--534, 1982.
\newblock \doi{10.1007/BF01457454}.

\bibitem[Thompson et~al.(2022)Thompson, Aktulga, Berger, Bolintineanu, Brown,
  Crozier, {in 't Veld}, Kohlmeyer, Moore, Nguyen, Shan, Stevens, Tranchida,
  Trott, and Plimpton]{lammps}
A.~P. Thompson, H.~M. Aktulga, R.~Berger, D.~S. Bolintineanu, W.~M. Brown,
  P.~S. Crozier, P.~J. {in 't Veld}, A.~Kohlmeyer, S.~G. Moore, T.~D. Nguyen,
  R.~Shan, M.~J. Stevens, J.~Tranchida, C.~Trott, and S.~J. Plimpton.
\newblock {LAMMPS} - a flexible simulation tool for particle-based materials
  modeling at the atomic, meso, and continuum scales.
\newblock \emph{Comput. Phys. Commun.}, 271:\penalty0 108171, 2022.
\newblock \doi{10.1016/j.cpc.2021.108171}.

\bibitem[Mishin et~al.(2001)Mishin, Mehl, Papaconstantopoulos, Voter, and
  Kress]{mishin2001}
Y.~Mishin, M.~J. Mehl, D.~A. Papaconstantopoulos, A.~F. Voter, and J.~D. Kress.
\newblock Structural stability and lattice defects in copper: Ab initio,
  tight-binding, and embedded-atom calculations.
\newblock \emph{Phys. Rev. B}, 63:\penalty0 224106, 2001.
\newblock \doi{10.1103/PhysRevB.63.224106}.

\bibitem[Sun et~al.(2006)Sun, Mendelev, Becker, Kudin, Haxhimali, Asta, Hoyt,
  Karma, and Srolovitz]{sun2006}
D.~Y. Sun, M.~I. Mendelev, C.~A. Becker, K.~Kudin, T.~Haxhimali, M.~Asta, J.~J.
  Hoyt, A.~Karma, and D.~J. Srolovitz.
\newblock Crystal-melt interfacial free energies in hcp metals: A molecular
  dynamics study of {Mg}.
\newblock \emph{Phys. Rev. B}, 73:\penalty0 024116, 2006.
\newblock \doi{10.1103/PhysRevB.73.024116}.

\bibitem[Mishin et~al.(2002)Mishin, Mehl, and Papaconstantopoulos]{mishin2002}
Y.~Mishin, M.~J. Mehl, and D.~A. Papaconstantopoulos.
\newblock Embedded-atom potential for {B2-NiAl}.
\newblock \emph{Phys. Rev. B}, 65:\penalty0 224114, 2002.
\newblock \doi{10.1103/PhysRevB.65.224114}.

\bibitem[Zhou et~al.(2004)Zhou, Johnson, and Wadley]{zhou2004eam}
X.~W. Zhou, R.~A. Johnson, and H.~N.~G. Wadley.
\newblock Misfit-energy-increasing dislocations in vapor-deposited {CoFe/NiFe}
  multilayers.
\newblock \emph{Phys. Rev. B}, 69:\penalty0 144113, 2004.
\newblock \doi{10.1103/PhysRevB.69.144113}.

\bibitem[Apostol and Mishin(2011)]{apostol2011}
F.~Apostol and Y.~Mishin.
\newblock Interatomic potential for the {Al-Cu} system.
\newblock \emph{Phys. Rev. B}, 83:\penalty0 054116, 2011.
\newblock \doi{10.1103/PhysRevB.83.054116}.

\bibitem[Hennig et~al.(2008)Hennig, Lenosky, Trinkle, Rudin, and
  Wilkins]{hennig2008}
R.~G. Hennig, T.~J. Lenosky, D.~R. Trinkle, S.~P. Rudin, and J.~W. Wilkins.
\newblock Classical potential describes martensitic phase transformations
  between the $\alpha$, $\beta$, and $\omega$ titanium phases.
\newblock \emph{Phys. Rev. B}, 78:\penalty0 054121, 2008.
\newblock \doi{10.1103/PhysRevB.78.054121}.

\bibitem[Andre et~al.(2021)Andre, Xie, Ott, P{\"u}rstl, Lohrey, Clegg,
  Sandl{\"o}bes-Haut, and Korte-Kerzel]{andre2021theta}
D.~Andre, Z.~Xie, F.~Ott, J.~T. P{\"u}rstl, N.~Lohrey, W.~J. Clegg,
  S.~Sandl{\"o}bes-Haut, and S.~Korte-Kerzel.
\newblock Dislocation-mediated plasticity in the {Al$_2$Cu} $\theta$-phase.
\newblock \emph{Acta Mater.}, 209:\penalty0 116748, 2021.
\newblock \doi{10.1016/j.actamat.2021.116748}.

\bibitem[Nos{\'e}(1984)]{nose1984}
S.~Nos{\'e}.
\newblock A unified formulation of the constant temperature molecular dynamics
  methods.
\newblock \emph{J. Chem. Phys.}, 81:\penalty0 511--519, 1984.
\newblock \doi{10.1063/1.447334}.

\bibitem[Hoover(1985)]{hoover1985}
W.~G. Hoover.
\newblock Canonical dynamics: Equilibrium phase-space distributions.
\newblock \emph{Phys. Rev. A}, 31:\penalty0 1695--1697, 1985.
\newblock \doi{10.1103/PhysRevA.31.1695}.

\bibitem[Leipold et~al.(2016)Leipold, Lazar, Brakke, and
  Srolovitz]{leipold2016}
H.~Leipold, E.~A. Lazar, K.~A. Brakke, and D.~J. Srolovitz.
\newblock Statistical topology of perturbed two-dimensional lattices.
\newblock \emph{J. Stat. Mech.}, page 043103, 2016.
\newblock \doi{10.1088/1742-5468/2016/04/043103}.

\bibitem[Lazar and Shoan(2020)]{lazar2020chains}
E.~A. Lazar and A.~Shoan.
\newblock Voronoi chains, blocks, and clusters in perturbed square lattices.
\newblock \emph{J. Stat. Mech.}, page 103204, 2020.
\newblock \doi{10.1088/1742-5468/abb6e3}.

\end{thebibliography}

\clearpage
\setcounter{figure}{0}\renewcommand{\thefigure}{S\arabic{figure}}
\setcounter{table}{0}\renewcommand{\thetable}{S\arabic{table}}
\setcounter{equation}{0}\renewcommand{\theequation}{S\arabic{equation}}
\setcounter{section}{0}\renewcommand{\thesection}{S\arabic{section}}
\renewcommand{\theHsection}{\thesection}\renewcommand{\theHfigure}{S\arabic{figure}}\renewcommand{\theHtable}{S\arabic{table}}\renewcommand{\theHequation}{S\arabic{equation}}
\begin{center}{\LARGE Supplementary Information}\\[6pt]{\large Burgers vectors from Voronoi topology}\end{center}
\section{Algorithmic details}\label{sec:si-alg}

\subsection{Frame relations in two dimensions}
\begin{figure}[!htbp]
\centering
\includegraphics[width=0.40\textwidth]{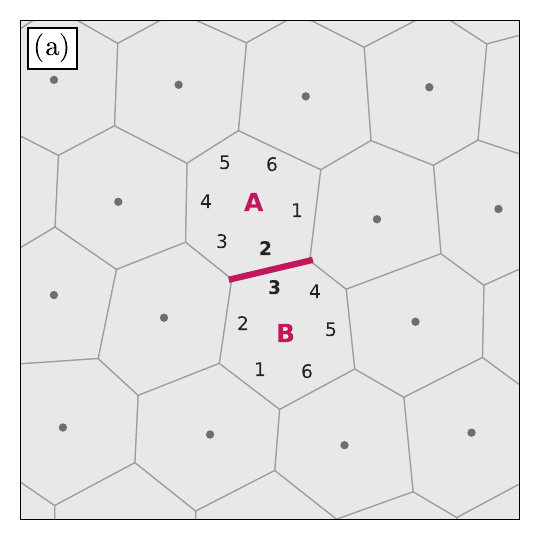}\hspace{0.04\textwidth}%
\includegraphics[width=0.40\textwidth]{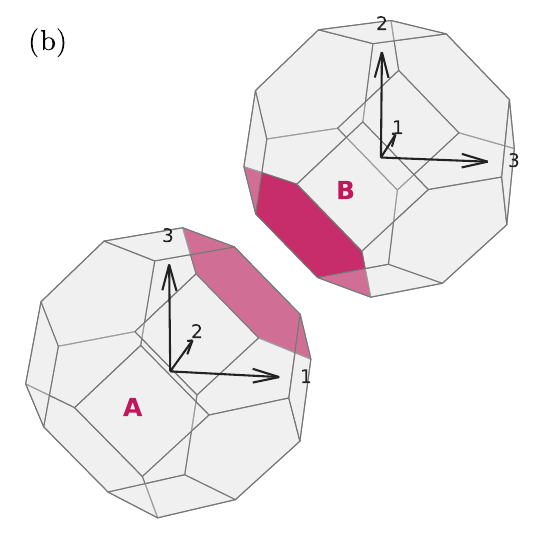}
\caption{\textbf{Frame relations in two and three dimensions.} (a) Two neighboring hexagons $A$ and $B$ in a triangular crystal with
thermal noise number their sides from arbitrary starts. Their shared side (red) is side $k=2$ of $A$ and side $l=3$ of $B$,
so adding $k+3-l=2$ (mod~6) to the side numbers of $B$ gives those of $A$.
(b) The same in three dimensions, for two neighboring cells of a bcc crystal, truncated octahedra drawn apart. Each cell's frame is three numbered axes, and the two cells number the same directions differently. Their shared
hexagonal face (red) fixes the crystal symmetry relating the two frames, here a third of a turn about the line
through the atoms.}
\label{fig:si-frames2d}
\end{figure}

In a triangular crystal, each hexagon numbers its six sides from 1 to 6, starting from any side and turning in the
same sense in every cell (clockwise in Fig.~\ref{fig:si-frames2d}a). This numbering is the cell's \emph{frame}. The
neighbor across side~$k$ lies in direction~$k$, so the frame also numbers the six directions to the neighbors. Suppose
cells $A$ and $B$ share a side that $A$ numbers $k$ and $B$ numbers $l$. Then direction $l$ of $B$ points back along
direction $k$ of $A$, so it is $A$'s direction $k+3$, and adding $k+3-l$ (mod~6) to the side numbers of $B$ gives
those of $A$. This rotation, together with the ideal step in direction~$k$ of $A$, is the frame relation of the two
cells. Supplementary Movie~1 follows these steps around a whole circuit. The three-dimensional frame relations
generalize this rule: faces take the place of sides, and crystal symmetries take the place of rotations by sixths of a
turn (Fig.~\ref{fig:si-frames2d}b).

\subsection{The Voronoi stability gap}
\begin{figure}[!htbp]
\centering
\includegraphics[width=0.6\textwidth]{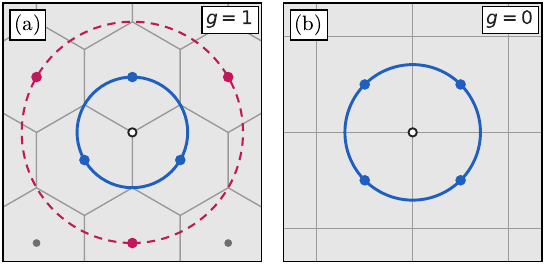}
\caption{\textbf{The Voronoi stability gap in two dimensions.} Each panel shows a Voronoi vertex (open circle), its
nearest atoms (blue), and the circle through them. (a) Triangular crystal: three atoms lie on the circle, and the next
ones (red) are twice as far, so $g=1$. (b) Square crystal: four atoms lie on the circle, so four cells meet at the
vertex and $g=0$. An arbitrarily small displacement changes which cells meet there.}
\label{fig:si-gap}
\end{figure}

Figure~\ref{fig:si-gap} illustrates the stability gap $g$ (Methods) in two dimensions, where each Voronoi vertex is
the center of a circle through three or more atoms, with none inside, and $r_3$ and $r_4$ take the place of $r_4$ and
$r_5$. The triangular crystal has $g=1$. The square crystal, whose vertices are each shared by four cells
(degenerate), has $g=0$.

\subsection{Why the vectors are exact}\label{sec:si-exact}
In a topologically stable crystal without topological pseudo-symmetry, four facts make the Burgers vectors exact. All
four rest on one elementary fact: a crystal symmetry, like any isometry, is fixed by where it sends four points not all
in one plane, so the only crystal symmetry that fixes four such points is the identity.

\emph{Frame relations are unique.} Let cells $A$ and $B$ share a face. Each edge of that face is shared with a third
cell, whose atom is a neighbor of both $A$ and $B$, so both labelings give it an ideal position. These atoms, at least
three, surround the line through $A$ and $B$, so together with $A$ and $B$ they include four points not all in one
plane. At most one crystal symmetry therefore carries the positions given by $A$ onto those given by $B$. Once both
cells are labeled, their frame relation is fixed, and the frame relation from $B$ to $A$ is its inverse.

\emph{Loops around a Voronoi edge close.} Let crystalline cells $u$, $v$, and $w$ share a Voronoi edge, whose two ends
are vertices shared with a fourth cell, $x$ at one end and $y$ at the other. In a topologically stable crystal, any two
of the four cells at a vertex share a face, so each of $u$, $v$, and $w$ gives ideal positions to all five atoms. Each
frame relation carries one cell's positions of these atoms onto the next cell's, so the composition of the three
around the edge carries the positions given by $u$ back onto themselves. Since $x$ and $y$ lie on opposite sides of the
plane through $u$, $v$, and $w$, the five atoms are not all in one plane, and the composition is the identity: no
rotation and no translation. The argument treats the five atoms as distinct. In a periodic box shorter than about
three neighbor distances, a triangle of neighbors can instead wrap around the box, and the software reports such
loops (Methods).

\emph{A loop's vector depends only on how it winds.} A step from one cell to a neighbor and straight back contributes
nothing, since the two frame relations are inverses. Moving a loop across a Voronoi edge whose three cells are
crystalline inserts or removes the smallest loop around that edge, which also contributes nothing. Two loops that can
be deformed into each other through crystalline cells therefore return the same vector, and a loop that can be shrunk
to a point returns zero. When the net rotations are the identity, as around dislocations, the net translation of two
loops joined at a common cell is the sum of their net translations. On the boundary of a defect cluster, a closed
surface with $h$ handles, every closed curve can be deformed into an integer combination of $2h$ basic curves, two for
each handle~\cite{eppstein2003}, so their vectors determine those of all the others.

\emph{Vectors add at junctions.} Where dislocation lines meet, a curve around one branch can be deformed on the
boundary into the sum of curves around the other branches, all oriented the same way along the lines. By the
additivity above, its vector is the sum of theirs, so Burgers vector is conserved without being imposed.

Far from defects, without pseudo-symmetry, every labeling is the true one up to one symmetry for the whole region. A
loop that can be enlarged through crystalline cells into such crystal therefore returns the true Burgers vector.

\subsection{Implementation details}
\subsubsection{Reference cells}
Reference cells (Methods) are taken from the Voronoi tessellation of a large finite cluster of the ideal crystal, from
one atom near its center for each Wyckoff site. spglib also gives the point group $\mathcal G$, so the input need not
declare its symmetry, and its coordinates may be rounded. Relative to that atom, each face $F$ stores the ideal vector
$\mathbf d_F$ to the neighbor across it, and each vertex $V$ its ideal position $\mathbf P_V$, the center of the
sphere through the four atoms whose cells meet there.

For each site we compare two counts: $n_\mathrm{site}$, the number of elements of $\mathcal G$ that map the atom's
neighbors, with their species, onto themselves, and $n_\mathrm{comb}$, the number of combinatorial isomorphisms of the
reference cell onto itself or onto its mirror image (all face cycles reversed). Each site symmetry gives such an
isomorphism, so $n_\mathrm{site}\le n_\mathrm{comb}$. We call a site \emph{rigid} if the two counts are equal and no
other site has a cell with the same number of faces with each number of edges. Any two labelings of a cell onto the
reference of a rigid site then differ by a site symmetry, so the cell has only one labeling, up to symmetry. A site that
is not rigid does not imply pseudo-symmetry: its cell alone allows several labelings, but its neighbors usually leave
one (Section~\ref{sec:si-fixing}).

\subsubsection{Labelings}
A cell with a vertex shared by more than four cells receives no labeling here (degenerate cells are treated in
Section~\ref{sec:si-degenerate}). Each sample cell is compared with every reference cell and its mirror image, and each combinatorial isomorphism found
is a labeling of the cell. To save time, only references with the same number of faces with each number of edges are
tried. Isomorphisms are found by a standard
search: one directed edge of the cell is sent to each directed edge of the reference in turn, and the map is extended
face by face~\cite{weinberg1966}. If species are given, the atom must have the species of the reference site. For a
rigid site, all labelings are equivalent, so the search stops at the first one. For any other site, each face of the reference is also tagged with
the number of faces with each number of edges of the neighbor's cell, and with the neighbor's species if species are given, and an
isomorphism must preserve these tags. This removes some isomorphisms that are symmetries of the polyhedron but not of the atom's surroundings in
the crystal. If the tags leave only site symmetries, one labeling again suffices. Otherwise all labelings are kept, and
agreement with neighbors removes the rest (Section~\ref{sec:si-fixing}). Through a labeling, each face $f$ of a sample
cell $i$ inherits an ideal vector $\mathbf d_i(f)$, and each vertex $v$ an ideal position $\mathbf P_i(v)$. We write
$\mathbf d_i(j)$ for $\mathbf d_i(f)$ when $f$ is the face that cell $i$ shares with cell $j$.

\subsubsection{Frame relations}
To find the frame relation $(R,\mathbf d)$ of Eq.~\eqref{eq:gluing} (Methods), with $\mathbf d=\mathbf d_i(j)$, we
solve that equation at three vertices of the shared face, take the nearest element of $\mathcal G$, and accept it only
if the equation holds at every vertex, in both directions. Only ideal positions enter, so the tolerance needs only to
absorb the rounding of the input: $10^{-6}$ of $(V/N)^{1/3}$, the cube root of the average volume per atom, for a
structure snapped onto its space group (Methods), and 0.2\% otherwise.

\subsubsection{Fixing the labelings}\label{sec:si-fixing}
Arc consistency, seeds, waiting cells, and the misfit are as in Methods. Arc consistency runs in rounds, each
deciding every cell from the labelings left at its start, so of two neighbors that cannot both be right, neither is
kept, and the result does not depend on the order in which cells are processed. In a growing region, cells are taken in order of how many labeled neighbors
they have.

\emph{Topological pseudo-symmetry.} When the Voronoi tessellation of the crystal is more symmetric than the crystal
(Results), the seed has several inequivalent labelings that each extend over the region, and topology cannot decide
between them. The region is then grown from each, and we keep the one with the smallest mean misfit over the
region's cells.

\begin{figure}[!htbp]
\centering
\includegraphics[width=\textwidth]{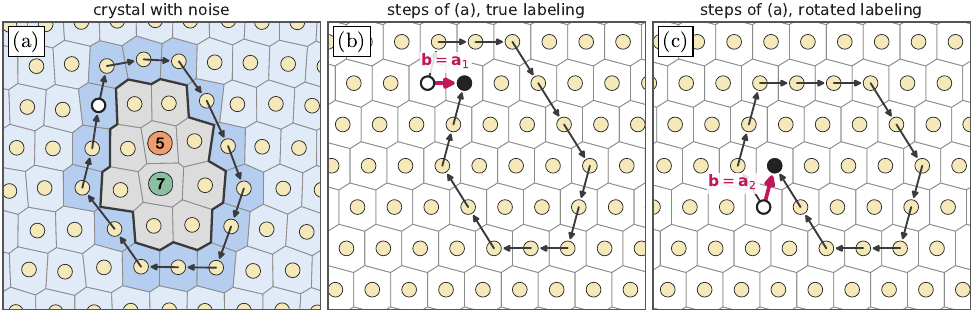}
\caption{\textbf{Topology alone cannot determine the Burgers vector in an oblique crystal.} All cells are hexagons, so
rotating every labeling by one face keeps them consistent. (a) An edge dislocation in a crystal with noise: its defect
cluster (gray), shell (blue), and circuit, from the open circle. (b) The circuit's steps read with the true labeling,
drawn in the ideal crystal. Their sum, the red arrow from the open circle to the filled one, is $\mathbf b=\mathbf a_1$.
(c) Read with the rotated labeling, they give $\mathbf b=\mathbf a_2$, also a lattice vector. Figs.~\ref{fig:concept} and~\ref{fig:gallery} show the
complementary picture: a circuit that is open in the sample but would close in the ideal crystal, with the red step
closing it in the sample.}
\label{fig:si-oblique}
\end{figure}

The simplest example is two-dimensional (Fig.~\ref{fig:si-oblique}). Every Voronoi cell of a generic oblique lattice is a hexagon, so its
tessellation has the topology of a honeycomb. That topology has 12 symmetries, six times the 2 of the oblique lattice.
Labeling one cell determines the labelings of all others, so a region has six labelings. They differ by which actual
neighbor vector each ideal one is assigned to. For example, the labeling rotated by one face reads each step $\mathbf a_1$ as $\mathbf a_2$, and
$\mathbf a_2$ as $\mathbf a_2-\mathbf a_1$. A loop then returns a lattice vector, but not the right one
(Fig.~\ref{fig:si-oblique}).
Geometry then tells the labelings apart, since only the true one maps the ideal neighbor vectors close to the actual ones. How clearly it does so depends on the lattice, and here pseudo-symmetry and the gap pull
in opposite directions. A lattice close to rectangular has a short pair of sides and a small gap, but its labelings are
easily told apart. A lattice close to triangular has a large gap, but its labelings fit almost equally well, and at
exactly triangular they become equivalent. Without geometry, only a known orientation of the crystal could tell them
apart. The space group, for example from a CIF file, does not help: it constrains only the frame relations, and these
are the same in every labeling. In three dimensions the same happens in a slightly sheared bcc crystal, whose cells are
still truncated octahedra.

Pseudo-symmetry is a property of the ideal crystal and can be detected from it alone, by counting the inequivalent
labelings of a crystalline region and comparing their misfits. We did this for all 9{,}610 structures of the Materials
Project test, each snapped onto its declared space group as in the test (all but 9, Methods). Each sample was a periodic block of
$n\times n\times n$ unit cells without noise, with $n\ge4$ chosen to give at least about 600 atoms. In every sample
all atoms were crystalline and formed a single region. In 9{,}220 structures (95.9\%) the seed has a single labeling
up to symmetry. In the other 390 (4.1\%) it has several, most often two to four. These counts are upper bounds, because arc consistency, which we use to list the
labelings, also admits some that do not extend over the whole region. Without noise the best labeling fits exactly.
The runner-up has a misfit below 0.01~\AA\ in 3 structures, 0.01--0.1~\AA\ in 70, 0.1--0.3~\AA\ in 128, and above
0.3~\AA\ in 189. For each sample, the software reports the smallest margin, in misfit, between the best labeling and
the runner-up over its regions, and warns when it is below 0.01~\AA.

The one structure in which the misfit had to decide a cell (Methods), CaCeHg$_2$, was analyzed as given,
because no tolerance reproduces its declared space group (Supplementary Section~\ref{sec:si-mp}). Its reference
therefore has a lower symmetry than the declared one, and atoms that the declared symmetry makes equivalent have nearly
identical cells, which the frame relations cannot tell apart.

\subsubsection{Loops and their net translations}
Loops, their net rotation and net translation $(R,\mathbf t)$, and the sign convention for $\mathbf b=\mathbf t$ are as
in Methods. For every Voronoi edge whose three cells are crystalline, the software checks that the loop around it
closes, as it must in topologically stable crystals (Results).

Each cell's frame is related to that of its region's seed by composing frame relations along any path of crystalline
cells. When no loop of the region has a net rotation, as only a disclination would, the result does not depend on
the path. All Burgers vectors of a region are thus expressed in one crystal frame, that of the seed up to a symmetry
of the crystal. Among these equivalent settings, the software reports the one whose axes lie closest to an orientation
supplied by the user, by default the laboratory axes.

\subsubsection{Boundary surfaces, loops, and per-atom vectors}
\begin{figure}[!htbp]
\centering
\includegraphics[width=0.9\textwidth]{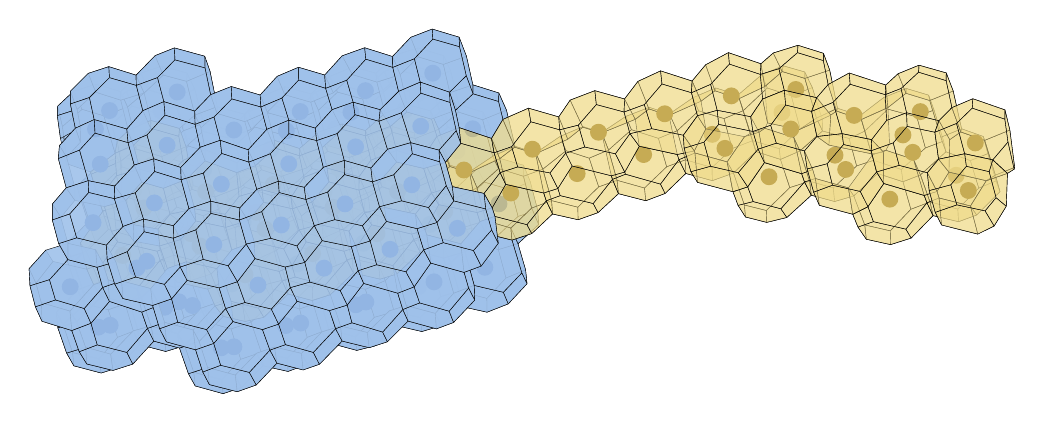}
\caption{\textbf{A defect cluster and its shell in three dimensions.} A segment of an edge dislocation in bcc Fe: the
Voronoi cells of its defect cluster (yellow) and, along the left half, its shell of crystalline cells (blue), each with the topology of the truncated octahedron, the Voronoi cell of bcc. The
boundary between them, made of Voronoi faces, is a tube around the dislocation.}
\label{fig:si-shell}
\end{figure}

As in Methods, a defect cluster is a connected set of atoms that are not crystalline or are treated as defects, and its boundary surface $\Sigma$
is made of the Voronoi faces it shares with the crystal (Fig.~\ref{fig:si-shell}). Each handle of $\Sigma$ carries two
closed curves, each curve becomes a walk through the shell atoms beside it, and the net translations of these walks
give the cluster's Burgers vectors. This section adds details of each step.

\emph{Surface.} In a configuration that is not generic, as in some samples without noise, $\Sigma$ need not be a closed
surface. The cluster is then flagged in the output.

\emph{Curves.} The $2h$ curves come from the tree--cotree construction~\cite{eppstein2003}. We take a spanning tree of
the faces, joining two faces that share an edge, and then a spanning tree of the vertices through the edges that the
first tree does not use. Exactly $2h$ edges are in neither tree. Each joins two faces, and with the path between them
in the first tree it forms one curve. A sphere has no such curves, so a cluster enclosed only by spheres, such as a
thermal fluctuation or a point defect, has no Burgers vector.

\emph{Walks.} Consecutive faces of a curve often border the same shell atom, so a walk can be shorter than its curve.
A walk that visits only one or two atoms returns the identity. For example, a ring-shaped cluster around a single
crystalline atom has a curve that runs around the inside of the ring, beside that atom alone. In close-packed crystals,
consecutive shell atoms that are partial neighbors, at opposite corners of a split octahedral hole, are related through
the hole's four other atoms. A step that still has no frame relation is replaced by a two-step detour through a common
full neighbor outside the cluster, and all such detours must give the same frame relation, or the curve is not used.

\emph{Content.} In close-packed crystals, regions grow only through full neighbors, while consecutive shell atoms may
be partial neighbors, so a component of $\Sigma$ could face more than one region. Its curves would then not be
combined, and the software reports it. None did in our samples. The integer combinations of Methods are found by row
operations on the curves' windings, and each is rounded to the nearest translation of the crystal.

\emph{Per-atom Burgers vectors.} The shortest walk through each shell atom (Methods) is found exactly by a
breadth-first search that tracks, for each atom reached, the translation accumulated on the way. This search is needed
only at some atoms. If atoms $x$ and $y$ are $k$ steps apart, a walk through $x$ becomes one through $y$ with the same
net translation and $2k$ more steps, by walking from $y$ to $x$ and back. So if at $x$ the shortest walk with one
vector is $m$ steps shorter than that with any other, every atom fewer than $m/4$ steps from $x$ has that vector alone.
The remaining atoms, mostly near junctions, are searched one by one.

\emph{Circuits for figures.} The circuits drawn in Fig.~\ref{fig:gallery} and Supplementary
Figs.~\ref{fig:si-gallery-screw}--\ref{fig:si-gallery-edge} (Methods) run once clockwise around the core through
crystalline atoms, from a start (open circle) to an end (filled circle). By convention, the end is an atom of the shell
walk that returned $\bvec$, which also fixes the sign of $\bvec$, and the start is the crystalline atom that lies
$\bvec$ from the end in the ideal crystal. Of all such circuits, over every choice of end, we draw a shortest one in
total length.

\subsubsection{Degenerate cells}\label{sec:si-degenerate}
\begin{figure}[!htbp]
\centering
\sbox0{\includegraphics[height=0.2\textwidth]{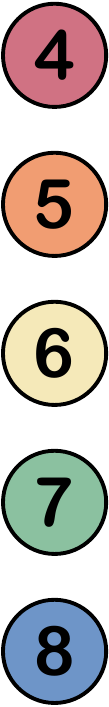}}%
\raisebox{-0.5\height}{\includegraphics[width=\dimexpr\textwidth-\wd0-0.8em\relax]{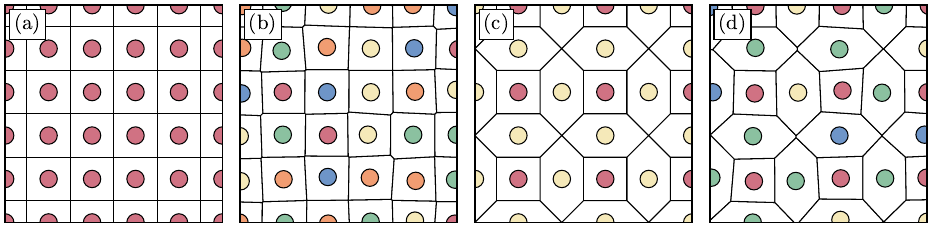}}\hspace{0.8em}%
\raisebox{-0.5\height}{\usebox0}
\caption{\textbf{Two degenerate crystals in the plane}, ideal and with Gaussian noise of $0.03\,d$ per component, each
atom colored by the number of sides of its cell (key at right). (a, b) The square crystal. Every vertex is shared by
four cells, and with noise each splits into two vertices joined by a short partial side. (c, d) The Lieb
crystal, with square and hexagonal cells. Only the vertices shared by four hexagons are degenerate, so the square cells
remain squares.}
\label{fig:si-degenerate2d}
\end{figure}

\emph{Why degenerate cells cannot be read one at a time.} In a degenerate crystal (such as fcc and hcp, but not bcc), a small displacement changes the
topology of the cells, so a cell in a finite-temperature sample need not have the ideal topology. The square crystal shows this in its
simplest form (Fig.~\ref{fig:si-degenerate2d}a, b). Each ideal cell is a square, and four cells meet at each corner. Its
full neighbors are the four atoms across its sides, and its partial neighbors the diagonal ones. A small
displacement splits each corner into two vertices joined by a short side, a \emph{partial side}, which makes one
diagonal pair of the four cells neighbors. A cell gains such a side at each corner where it belongs to that pair, so a
cell at finite temperature has four to eight sides, and its polygon does not tell which four are the sides of the ideal
square. In the limit of small noise only about
4.4\% of the cells remain squares~\cite{leipold2016}. The statistics of chains, blocks, and clusters of such cells
have also been derived~\cite{lazar2020chains}. Geometry separates the two kinds of sides: a partial side is about as long as the noise, and a side shared with a full neighbor
about $d$. A rule based on lengths would need a threshold, however, and we use topology instead. In
fcc and hcp the problem is worse. Each ideal cell has vertices where four of its own faces meet (a generic cell
has three faces at every vertex), so these vertices always split and no cell keeps the ideal topology. In our
samples with thermal-like noise only 5--7\% of the cells have exactly twelve faces, one per full neighbor, and many admit more than
one choice of full-neighbor faces.

\emph{A table of cell topologies.} The table of Methods is built as follows for fcc. The
fcc cell has six degenerate (octahedral) vertices. Seen from the central cell, each can split into an edge (in one of
two orientations), a triangular face (in one of four positions), or a quadrilateral face (in one way): three kinds of
splitting and seven splittings in all. The enumeration of these splittings over all six degenerate vertices gives
2{,}815 topologies for fcc, and the same enumeration gives 9{,}556 for hcp. Eight fcc and 223 hcp topologies have
more than one alignment. Read one cell at a time, the table
leaves more than one alignment for 10--12\% of the cells in our Cu snapshots and 62--64\% in Mg (all 84 snapshots of
Supplementary Section~\ref{sec:si-val}).

\emph{Neighbor agreement.} Away from defects the true
alignments of neighboring cells always agree, so agreement (Methods) never removes them, and a cell left with a single alignment
has its true full neighbors. In the Cu and Mg snapshots, agreement leaves more than one alignment for at most 0.06\% of
the Cu cells and 0.6\% of the Mg cells. Cells left with no alignment make up 0.5--1.0\%, and cells whose topology is
not in the table 0.2--1.3\%, rising with temperature. Near a dislocation, agreement may leave a strip of cells along
the glide plane undecided, because alternative alignments there place the core at different positions along the
plane. These cells join the defect cluster, which therefore extends along the glide plane. The Burgers vector,
returned by a circuit through the crystalline cells around the cluster, is unaffected.

\emph{Two-dimensional illustration.} The same construction applies in the plane. As in the main text, the topology of
a two-dimensional cell is its polygon together with the numbers of sides of its neighbors, in a canonical cyclic order~\cite{lazar2024twod}. An alignment is then a set of sides that join full neighbors.
In the square crystal, the topology of an atom's cell depends on the 16 vertices of the $3\times3$ block of cells centered on it,
all of them degenerate. All $2^{16}$ splittings of these vertices occur under small perturbations, and they give
1{,}261 topologies, 322 of which allow more than one alignment. The Lieb crystal (Fig.~\ref{fig:si-degenerate2d}c, d) has atoms at the corners and edge
midpoints of the squares of a square lattice. The centers of these squares, where no atom sits, are its only degenerate
vertices, each shared by the cells of four edge-midpoint atoms, and the same enumeration gives 55 topologies, none of
them ambiguous. The two crystals mirror fcc and hcp. In the Lieb crystal, as for most fcc cells, the table alone determines the full
neighbors of every cell. In the square crystal, as in hcp Mg, it leaves about two thirds of the cells with several
alignments, and agreement decides all of them. We tested both in periodic samples of $24\times24$ unit cells, five per
noise level, with Gaussian noise of $0.005a$--$0.05a$ per component, where $a$ is the side of the square unit cell ($a=d$ in the
square crystal and $2d$ in the Lieb crystal). In the square crystal, the table and agreement gave the true full
neighbors of every cell. In the Lieb crystal they did so for every cell up to $0.02a$. At $0.05a$, 3.5\% of the Lieb cells had also changed topology at generic vertices, so their topologies were not in the table and
they were treated as defects. In neither crystal was a cell given a wrong alignment.

\emph{Labeling by the full-neighbor arrangement.} In fcc and hcp, the twelve full neighbors of a cell with a single alignment, with
the full-neighbor bonds among them, are matched to their ideal arrangement in fcc or hcp stacking, written in the frame of the crystal, the hcp
one in both of its orientations, which differ by a half turn about the stacking axis. Matches must preserve orientation, taken from the oriented Delaunay tetrahedra.
Matching gives each full neighbor its ideal vector, and hence the cell's labeling. The hcp arrangement has more
combinatorial than geometric symmetry (its basal mirror), so it can admit a match that is
combinatorially allowed but geometrically wrong. Such arrangements are therefore labeled by consistency with their
labeled neighbors, starting from the cells with the most labeled neighbors and growing outward. Frame relations are
obtained from the common full neighbors $k$ of two cells $i$ and $j$, $\mathbf d_i(k)=R\,\mathbf d_j(k)+\mathbf
d_i(j)$, with $R$ in the point group of the crystal.

\emph{Stacking faults.} Stacking-fault atoms have the full-neighbor arrangement of the other stacking. Their topologies cannot exclude
them: 880 of the 2{,}815 fcc topologies also arise in hcp, among them the most common ones, so excluding these would
remove most of the crystal. Most stacking-fault atoms fail the checks (Methods), however, so a dissociated dislocation and its stacking fault form one cluster, with the Burgers vector of the perfect dislocation. Resolving each
partial would require recognizing stacking-fault atoms as crystal of the other stacking, with frame relations between the two,
which we leave to future work.

\emph{Wrong alignments near cores.} The checks compare arrangements, not distances. Near the core of a partial
dislocation the crystal is strongly sheared, and a partial neighbor, at $\sqrt2\,d$ in the ideal crystal, can come
almost as close as a full neighbor. A cell can then pass the checks with that partial neighbor in place of one of its
full neighbors. To measure how often, we quenched all 80 snapshots of the Cu runs of Supplementary Section~\ref{sec:md}
(16 each at 300, 600, 900, 1100, and 1200~K, including the two per temperature before 7~ps that Table~\ref{tab:md}
does not count) to 0~K by energy minimization. In the quenched positions, each atom's
twelve nearest atoms are its true full neighbors, and their arrangement gives its true stacking. None of the about
$2.3\times10^6$ atoms of fcc stacking that passed the checks had a wrong full neighbor. Of the stacking-fault atoms, about
6{,}600 per temperature, 12--17\% passed the checks, all with a wrong full neighbor, with no trend in temperature. So
did 637--1{,}511 of the 4{,}096 atoms per temperature at the partial cores, of neither stacking, fewer at higher
temperature. In all but one of these atoms exactly one of the twelve was wrong, a partial neighbor at a median
distance of $1.3\,d$ in the quenched positions. The wrongly aligned stacking-fault atoms lie mostly beside the cores: 53\% have
a core atom as a nearest neighbor, against 31\% of all stacking-fault atoms. These wrong alignments are still consistent with
one another and with the crystal around them. In all 80 snapshots every smallest loop, a triangle of crystalline atoms
that are full neighbors of one another, closed: $1.8\times10^7$ loops, $1.0\times10^5$ of them through a wrongly
aligned cell. Every Burgers vector was correct.

\emph{Flat faults and defect-free crystal.} Faults without dislocations gave no wrong alignments. In fcc and hcp
crystals of about 5{,}000 atoms with two flat stacking faults and Gaussian noise of $0.005$--$0.08$ times the
full-neighbor distance per component (21 samples of each), no fault atom passed the checks in fcc, and 11 of 21{,}168
did in hcp, all with their true full neighbors. No other atom had a wrong full neighbor. In defect-free periodic fcc
and hcp crystals of 4{,}000 atoms with Gaussian noise up to $0.12$ times the full-neighbor distance per component
(three samples per level), no atom passing the checks had a wrong full neighbor, and no cluster returned a Burgers vector. In
fcc every atom passed the checks up to $0.06$, and at least 99.5\% at $0.08$. In hcp at least 99.3\% passed
up to $0.06$, and 97.5--99.5\% at $0.08$.

\emph{The $\omega$ phase.} $\omega$-Ti (P6/mmm) has one atom on the 1a site and two on the 2d site, and is treated as
in Methods. With the potential of Hennig et al.~\cite{hennig2008}, its lattice constants are $a=4.606$~\AA\ and
$c=2.815$~\AA. The cell of a 2d atom has 11 faces of positive area and touches its six partial neighbors at single
points, while the cell of a 1a atom has 14 faces, which it keeps under perturbation. Combinations of splittings that
are compatible pair by pair need not close into a cell, so we keep only those in which every Delaunay triangle
containing the atom lies in exactly two Delaunay tetrahedra. This gives the 388 topologies~\cite{votypes2026a} and 399
labelings of Methods. The degenerate 2d cells of the exactly ideal crystal match none of them and are not crystalline.

The defect-free samples of Results had 3{,}360 atoms, with Gaussian noise of $0.005$, $0.01$, $0.02$, $0.03$, $0.04$,
and $0.05\,d$ per component, and their crystalline atoms formed a single region. The three dipoles were inserted in
supercells $28d$ wide, with nine noise realizations each, and 95--98\% of the atoms were crystalline. In molecular
dynamics, a $[0001]$ screw quadrupole ($\bvec=\mathbf c$) annihilated during equilibration with the protocol of the
other crystals, after about 18~ps at 300~K, hence the early snapshots of Methods. An independent count, the winding of
the atomic displacements around each line, found all four lines in every one of the 38 snapshots, in which 91.4--94.1\%
of the atoms were crystalline.

\section{Validation protocol and benchmarks}\label{sec:si-val}
\subsection{Materials Project test}\label{sec:si-mp}
\emph{Selection.} The gaps of Results were computed by examining every Voronoi vertex of each stored structure, so
they are exact up to the rounding of the stored coordinates, about $10^{-6}$. Of the 163{,}211 structures, 25\%, 11\%,
and 8\% have $g$ above 0.008, 0.05, and 0.1, respectively. The 93\% that match no fcc, hcp, bcc, or diamond
arrangement (Discussion) were found with spglib (tolerance 0.25~\AA), with all species replaced by one: 11{,}708
structures match one of these arrangements, and 104{,}187 of the other 151{,}503 have $10^{-6}<g<0.2$. Table~\ref{tab:mp-stats} lists the 9{,}610 structures with $g\ge0.2$ by crystal system. They comprise 128
elemental structures, 2{,}801 binary and 6{,}514 ternary compounds, and 167 compounds of four or five elements. The 2{,}302 that are
experimentally observed fall into 48 of the 58 topology families and 135 of the 242 geometries.

\begin{table}[!htbp]
\centering\footnotesize
\caption{\textbf{The 9{,}610 Materials Project structures of the test} ($g\ge0.2$), by crystal system: the numbers of
structures, of those observed experimentally, and of space groups, and the numbers of atoms and of Wyckoff sites in the
primitive cell as tested. All but 9 structures were snapped onto their declared space group (Methods).}
\label{tab:mp-stats}
\vspace{4pt}
\begin{tabular}{lrrrrrr}
\toprule
crystal system & structures & observed & space groups & atoms per cell & median atoms & Wyckoff sites\\
\midrule
cubic        & 6{,}556 & 1{,}211 & 16 & 1--156 & 4  & 1--20\\
tetragonal   &    663 &    270 & 25 & 1--34  & 4  & 1--14\\
orthorhombic &    903 &    133 & 29 & 1--100 & 8  & 1--32\\
hexagonal    &    817 &    518 & 10 & 3--96  & 12 & 1--30\\
trigonal     &    537 &    157 & 11 & 1--171 & 12 & 1--57\\
monoclinic   &    107 &     13 & 10 & 1--55  & 12 & 1--46\\
triclinic    &     27 &      0 &  2 & 4--55  & 18 & 3--55\\
\midrule
all          & 9{,}610 & 2{,}302 & 103 & 1--171 & 4 & 1--57\\
\bottomrule
\end{tabular}
\end{table}

\emph{Declared symmetry.} Snapping onto the declared space group (Methods) mostly removes rounding, but it matters
for the 34 structures whose files list a supercell of the declared crystal. Analyzed as given, such a structure asks
the topology to distinguish atoms that the declared symmetry makes equivalent. The Nb of Methods (mp-aaafzwer), seven
atoms in a triclinic cell, is within 0.1~\AA\ of the declared rhombohedral crystal (R$\bar3$m) with one atom per cell. As
given, 16\% of its atoms were not crystalline at noise $0.02\,d$, and one dipole was never clean. Snapped, every atom
was crystalline at the same noise, and every dipole was clean at $28d$. The 9 structures for which no tolerance up to
0.2~\AA\ reproduces the declared space group (Methods) are DyYIr$_2$, CaCeHg$_2$, two polymorphs of MgSi$_2$,
Ba$_2$SrNp, YbGdZn$_2$, SrMg, and two polymorphs of Rb.

\emph{Thermal stability.} Defect-free samples of the $28d$ screw supercell were analyzed with Gaussian noise of
$0.02\,d$ and $0.03\,d$ per component. At $0.02\,d$, at least 95\% of the atoms are crystalline in every
structure, and at least 99\% in 9{,}600. At $0.03\,d$, 9{,}500 structures reach 95\% and 9{,}323 reach 99\%. In no sample do the crystalline atoms form more than one
crystalline region.

\emph{Dislocation character.} The screw dipoles are pure screw by construction. The Burgers vector of the non-screw dipole
is the shortest lattice vector not parallel to the line $\mathbf s$, so it makes an angle of at least $60^\circ$ with
the line: at a smaller angle, $\bvec-\mathbf s$ or $\bvec+\mathbf s$ would be a shorter lattice vector, also not
parallel to $\mathbf s$. Hence the non-screw dipole is pure edge or mixed at $60$--$89^\circ$ to the line
(Results). Dislocations at about $45^\circ$, and lines along longer lattice vectors, are tested below.

\emph{Test.} Each dipole was inserted into a fully periodic, restricted-triclinic supercell, and the whole sample was
analyzed (Methods). We call a dipole \emph{clean} if its two lines come out as two separate defect clusters with $\pm\bvec$. At the first width, $28d$ (2{,}028 to 12{,}312 atoms, median 7{,}056; Results), the screw dipole was clean in
9{,}560 structures, the non-screw dipole in 9{,}281, and both in the 9{,}270 of Results. Of the 50 screw
and 329 non-screw dipoles that were not clean, 41 and 173 had a single cluster that returned $\pm\bvec$, typically because both cores
lay in one cluster whose boundary surface has a handle around each line. In the other 9 and 156, no cluster returned a Burgers vector, as when both cores lie in a single tube whose curves enclose both lines and return zero. These
dipoles were analyzed again in cross-sections $56d$ and $112d$ wide (8{,}640 to 49{,}248 and 42{,}336 to 165{,}888
atoms). At $56d$, 49 of these screw and 325 non-screw dipoles were clean, and at $112d$ the remaining
screw dipole and 4 non-screw dipoles were clean. Both were thus clean in all 9{,}610 structures.

The third dipole (Methods), edge or mixed at $60$--$90^\circ$ to $\bvec$, was clean in 9{,}468 structures at $28d$,
in 138 more at $56d$, and in the last 4 at $112d$. Counting the wider samples, the three dipoles required the
29{,}360 analyses of Results: 28{,}830 were clean, one per dipole, and none of the other 530 returned a wrong vector.
The samples were at least $6d$ long, which suffices in every structure tested (Methods). In a molecular crystal, where
$d$ is a short intramolecular bond, $6d$ may be too short.

\emph{Noise or strain?} A dipole may need a wider sample because the noise destroys the ideal cell topology near its
cores, or because the elastic strain does. To separate the two, the 340 structures whose screw or non-screw dipole needed
a wider sample were analyzed again with noise of $0.01\,d$ and $0.005\,d$, the samples being otherwise unchanged. Of
their 680 dipoles, 301 were clean at $28d$ with noise $0.02\,d$, 462 with $0.01\,d$, and 502 with $0.005\,d$. Noise
thus accounts for part of the widening. Even at a quarter of the noise, however, 178 of these dipoles (26\%) still
needed wider samples, which suggests that strain sets their topological core: it extends to a distance of order
$|\bvec|/(2\pi g)$ from the line, large when $g$ is small (Discussion). At every noise level every dipole was clean at
some width, and none returned a wrong vector.

\emph{Further Burgers vectors and lines.} For one structure of each of the 242 geometries, we took up to three Burgers
vectors, the shortest lattice vector of each distinct length up to twice the shortest ($1.0$--$11.3\,d$). Each was
inserted along up to three lines: along $\bvec$ (screw), along the short lattice vector closest to perpendicular to
$\bvec$, and along the one closest to $45^\circ$ from it. Each combination of vector and line was inserted as a dipole
in a fully periodic cell whose third vector is the line, with the protocol of the main test: $28d$ wide, then $56d$
and $112d$ while not clean, at least $6d$ long, and noise $0.02\,d$. For 69 of the 2{,}127 combinations, 66 of them
screw, no such cell could be built from short lattice vectors. Each of the other 2{,}058 returned the inserted vector
and no other. In 2{,}055 the two lines formed separate clusters, at $28d$ (1{,}585), $56d$ (398), or $112d$ (72). In
the other three they still shared one cluster at $112d$, which returned $\pm\bvec$. In two samples voro++ could not
tessellate a very thin, strongly tilted box, and the same sample, twice as long along the line, was used instead.

\subsection{Gallery of Burgers circuits}\label{sec:si-gallery}
\begin{figure}[p]
\centering
\includegraphics[width=\textwidth]{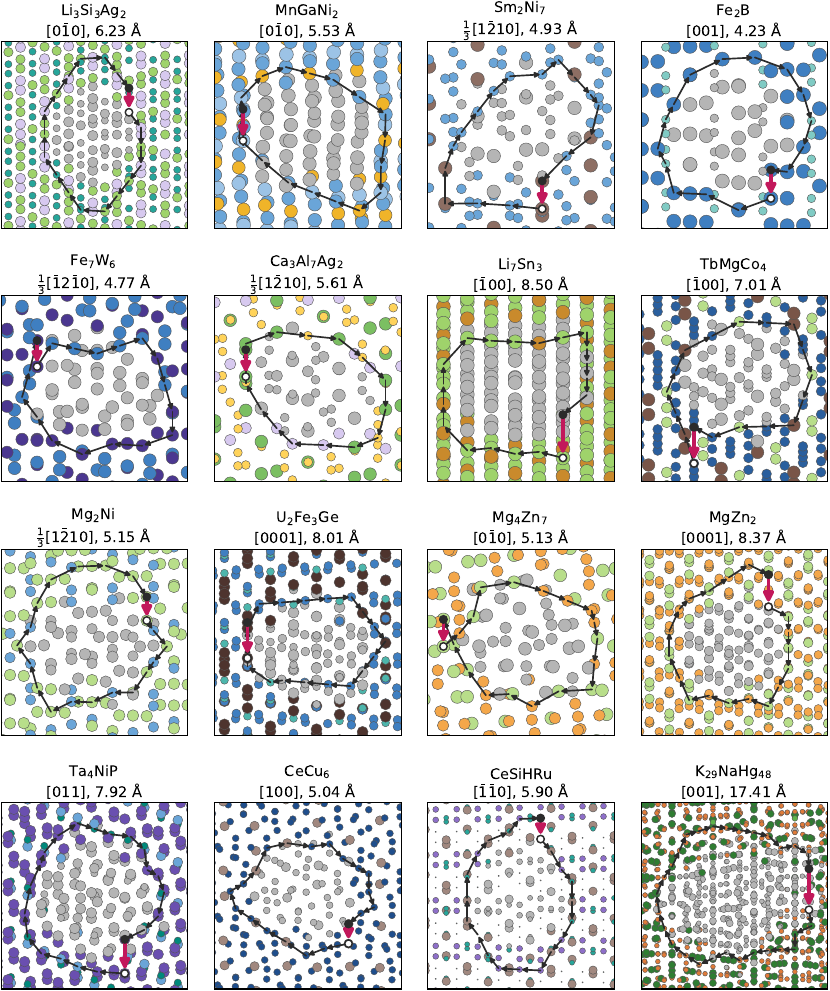}
\caption{\textbf{Screw dislocations in 16 crystals}, viewed $30^\circ$ off the line, drawn as in Fig.~\ref{fig:gallery}: atoms of the defect cluster are gray, and the others are colored
by element.}
\label{fig:si-gallery-screw}
\end{figure}

\begin{figure}[p]
\centering
\includegraphics[width=\textwidth]{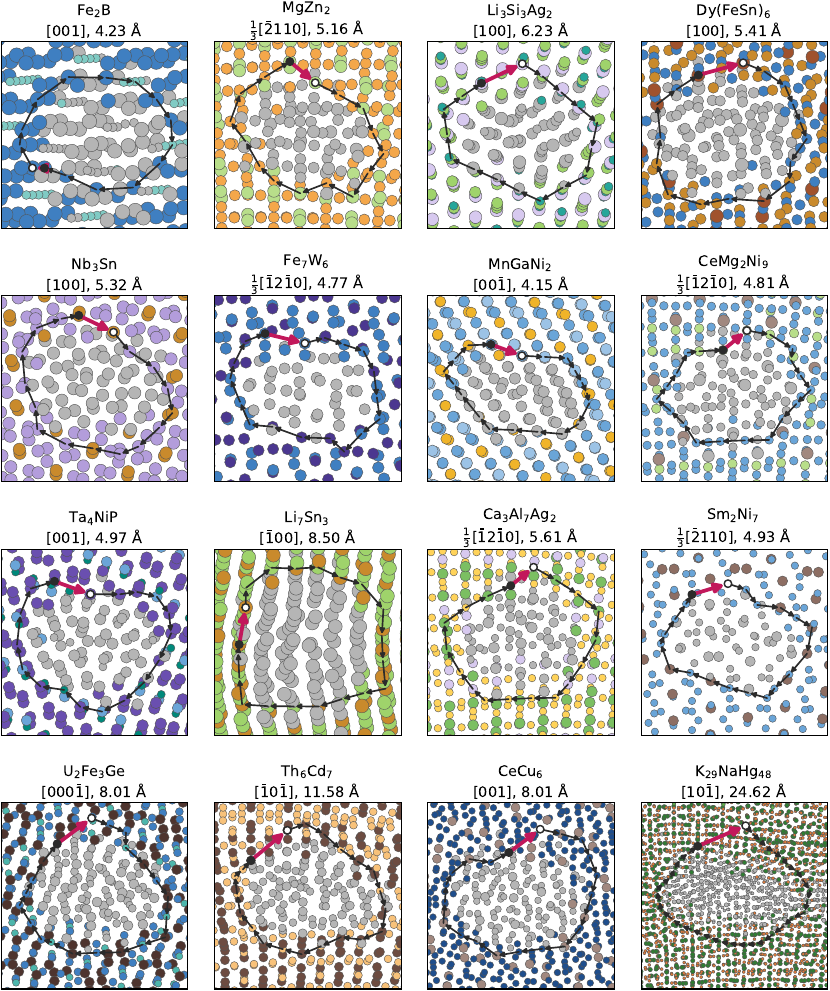}
\caption{\textbf{Mixed dislocations in 16 crystals} ($30$--$60^\circ$ between $\bvec$ and the line), viewed $30^\circ$
off the line, drawn as in Fig.~\ref{fig:gallery}: atoms of the defect cluster are gray, and the others are colored
by element.}
\label{fig:si-gallery-mixed}
\end{figure}

\begin{figure}[p]
\centering
\includegraphics[width=\textwidth]{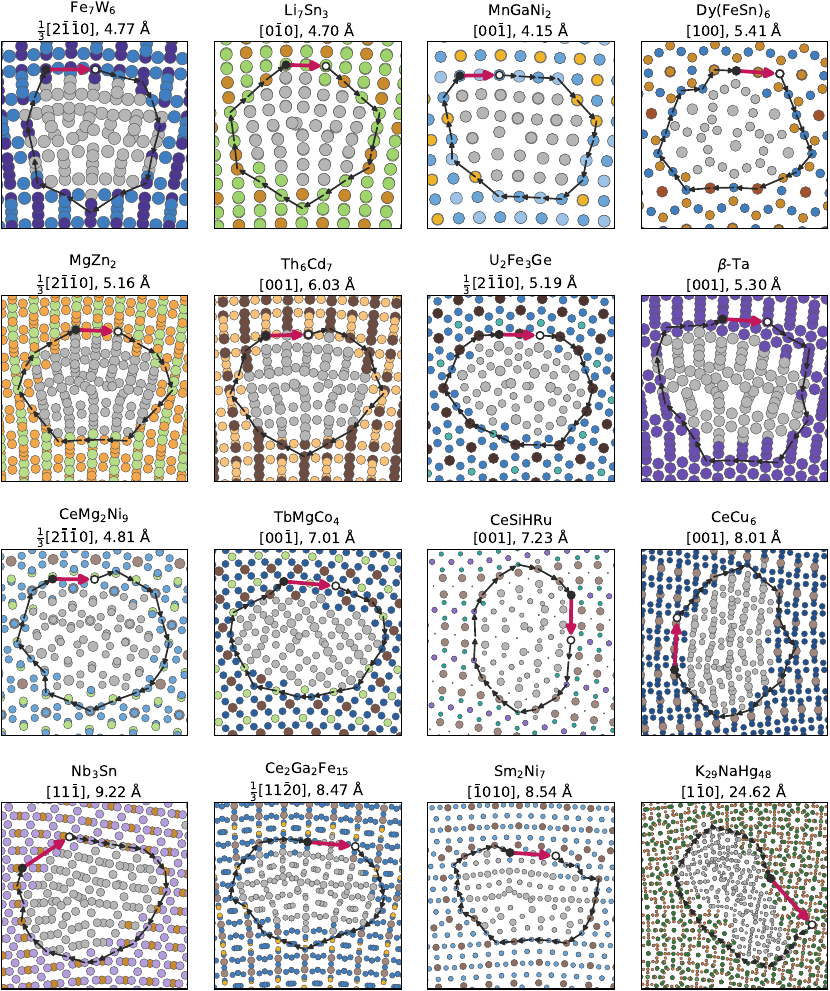}
\caption{\textbf{Edge dislocations in 16 crystals}, viewed along the line, drawn as in Fig.~\ref{fig:gallery}: atoms of the defect cluster are gray, and the others are colored
by element.}
\label{fig:si-gallery-edge}
\end{figure}

Supplementary Figs.~\ref{fig:si-gallery-screw}--\ref{fig:si-gallery-edge} show 16 screw, 16 mixed, and 16 edge
dislocations in crystals of the Materials Project test, drawn as in Fig.~\ref{fig:gallery}. Within each figure every
panel shows a different crystal, and none of them appears in the matching row of Fig.~\ref{fig:gallery}. These
dislocations are not those of the test itself. For each crystal, we took as Burgers vectors the shortest lattice
vector of each distinct length up to twice the shortest, and as lines short lattice vectors along $\bvec$ (screw), at
$30$--$60^\circ$ to it (mixed), or at $75^\circ$ or more (edge; $90^\circ$ in these figures). Each candidate was
inserted as a dipole and analyzed as in the test ($28d$ samples, widened while not clean, noise $0.02\,d$), and every
one returned its inserted vector. For each crystal we show the candidate whose circuit is roundest and smoothest seen
along the line, and within a figure the panels are ordered by the width of the core.

\subsection{Molecular-dynamics validation}
\label{sec:md}
This section gives in full the protocol summarized in Methods.

\emph{Samples.} Every sample is a fully periodic box, and the analysis receives only the atoms and the box. The supercell is spanned by three lattice vectors, with the lines along
the third, and is written in restricted-triclinic form. Each sample holds four lines with zero total Burgers vector,
so that the box is not sheared. The displacement is the same elastic field as in the Materials
Project test (Methods), summed over periodic images, superposed for the four lines, and periodic to within
0.002~\AA. Screw dislocations
form a checkerboard, with $+\bvec$ at $(0,0)$ and $(\tfrac12,\tfrac12)$ and
$-\bvec$ at $(\tfrac12,0)$ and $(0,\tfrac12)$, in fractions of the two cell vectors that span the cross-section. Edge and mixed dislocations form the standard $45^\circ$ quadrupole,
with $+\bvec$ at $(0,0)$ and $(0,\tfrac12)$ and $-\bvec$ at $(\tfrac12,\tfrac14)$ and $(\tfrac12,\tfrac34)$. Here the
second coordinate runs along the glide-plane normal, and the cell is doubled in that direction, so that partners of
opposite sign lie on different glide planes. Every line lies midway between two atomic glide planes. A cut between
lines on different glide planes has a climb component and doubles a slab of atoms. The extra atoms are removed by a
rule decided on the ideal crystal: atom $i+\bvec$ is removed when the segment from $i$ to $i+\bvec$ crosses the cut.
This removes exactly the slab volume times the density of each species (in the C15 edge sample, 288 Mg and 576 Cu
atoms), which a rule based on distances after displacement fails to do near cores with large $|\bvec|$ and two
species. The prismatic vacancy loop in Fe ($\bvec=\tfrac12[111]$, radius 20~\AA) was built by removing a disc of
three $(111)$ layers and closing the gap with the solid-angle displacement field.

The cells must be wide enough that the lines neither annihilate nor merge during the run. In narrower cells the screw quadrupoles of Fe and NiAl annihilated at 900~K and above,
and the $\theta$ lines shared clusters, so these cells of Table~\ref{tab:md} are twice
as wide in both directions of the cross-section. In Mg, $\langle a\rangle$ screw quadrupoles annihilated in every
arrangement we tried, in cells up to 200~\AA\ wide. The winding of the atomic $z$ coordinates around each line
confirms that the lines are physically gone, and the method correctly reports nothing, so Table~\ref{tab:md} has no
$\langle a\rangle$ result for Mg.

\emph{Molecular dynamics.} Samples were minimized with the box relaxed, kept orthogonal for Fe, NiAl, Cu, and Mg and
allowed to tilt for $\theta$ and C15. The close-packed analysis requires an orthogonal box, which is why Cu and Mg
keep theirs. Each sample was then run for 15~ps at constant temperature and zero pressure (NPT), and then, with the box fixed at
its average over the last 5~ps, at constant temperature (NVT), with a time step of 1~fs. After 5~ps, raw snapshots were taken every 1~ps (1.5~ps for
$\theta$) and analyzed without quenching or averaging, in Mg every third one and in Cu the 14 from 7~ps on. The
$\omega$-Ti samples followed a different protocol (Supplementary Section~\ref{sec:si-degenerate}). The minimized 0~K
configurations of the crystals with stable cells were also analyzed, and every one was resolved. They are not counted
in Table~\ref{tab:md}. Minimized Cu, Mg, and $\omega$-Ti configurations are not counted, because the cells of a minimized degenerate
crystal are exactly degenerate, outside the generic geometry the construction assumes.

\emph{Classification.} Each snapshot is placed in the first of four classes that applies. It is \emph{wrong} if any cluster returns a vector that is not an integer multiple of the inserted $\bvec$. It is \emph{resolved} if each
inserted line is a defect cluster of its own with exactly $\pm\bvec$ (in Cu, a dissociated dislocation and its
fault form one cluster, Section~\ref{sec:si-degenerate}). It is \emph{merged} if some cluster returns more than one vector, counting $+\bvec$ and
$-\bvec$ as different, or a multiple of $\bvec$ other than $\pm\bvec$, and so must contain several lines. It is
\emph{lost} if fewer clusters return a vector than lines were inserted. The classes are read from the vectors alone:
a cluster that holds two lines with the same vector returns only $\pm\bvec$, and counts as one line.

\begin{table}[!htbp]
\centering\footnotesize\setlength{\tabcolsep}{3.5pt}
\begin{tabular}{lllrrrrrr}
\toprule
crystal & dislocations & $T$ (K) & atoms & snaps. & resolved & merged & lost & wrong\\
\midrule
bcc Fe & edge $\tfrac12[111](\bar110)$, $59{\times}121$~\AA & 300, 900, 1200 & 16{,}920 & 63 & 63 & 0 & 0 & 0\\
 & edge, $119{\times}242$~\AA & 300, 900, 1200 & 68{,}400 & 63 & 63 & 0 & 0 & 0\\
 & & 1500 & 68{,}400 & 21 & 20 & 1 & 0 & 0\\
 & & 1650 & 68{,}400 & 21 & 0 & 21 & 0 & 0\\
 & & 1700, 1755 & 68{,}400 & 42 & 0 & 42 & 0 & 0\\
 & screw $\tfrac12[111]$, $168{\times}170$~\AA & 300, 900, 1200 & 96{,}768 & 63 & 63 & 0 & 0 & 0\\
 & & 1500 & 96{,}768 & 21 & 20 & 1 & 0 & 0\\
 & & 1650 & 96{,}768 & 21 & 0 & 21 & 0 & 0\\
 & & 1700, 1755 & 96{,}768 & 42 & 0 & 42 & 0 & 0\\
 & prismatic vacancy loop & 900, 1200 & 45{,}815 & 42 & 42 & 0 & 0 & 0\\
B2 NiAl & $[111]$ screw, B2 reference & 300, 900 & 96{,}768 & 42 & 42 & 0 & 0 & 0\\
C15 MgCu$_2$ & screw $\tfrac12[1\bar10]$ & 300 & 15{,}120 & 21 & 21 & 0 & 0 & 0\\
 & edge $\tfrac12[1\bar10](111)$ & 300 & 30{,}240 & 21 & 21 & 0 & 0 & 0\\
 & $60^\circ$ mixed $\tfrac12[10\bar1](111)$ & 300 & 33{,}696 & 21 & 21 & 0 & 0 & 0\\
$\theta$-Al$_2$Cu & 11 ($\bvec$, line) combinations & 300 & 32--81k & 110 & 110 & 0 & 0 & 0\\
 & & 500 & 32--81k & 110 & 95 & 15 & 0 & 0\\
fcc Cu & edge $\tfrac12[1\bar10](111)$, $102{\times}200$~\AA & 300, 600, 900, & 30{,}336 & 70 & 70 & 0 & 0 & 0\\
 & & 1100, 1200 & & & & & & \\
hcp Mg & screw $[0001]$, $192{\times}188$~\AA & 300, 600 & 48{,}960 & 14 & 14 & 0 & 0 & 0\\
$\omega$-Ti & screw $[0001]$, $193{\times}191$~\AA & 300, 600 & 60{,}480 & 38 & 38 & 0 & 0 & 0\\
\midrule
total & & & & 846 & 703 & 143 & 0 & 0\\
\bottomrule
\end{tabular}
\caption{Molecular-dynamics validation in fully periodic samples, analyzed whole, with each raw snapshot counted
once. Cu and Mg were analyzed with the table of cell topologies and neighbor agreement, and $\omega$-Ti with its
enumerated reference cells (Supplementary Section~\ref{sec:si-degenerate}).}
\label{tab:md}
\end{table}

\emph{Results.} No vector was wrong in any of the 846 snapshots, and every snapshot was resolved except 15 of $\theta$
at 500~K and 128 of Fe at 1500--1755~K, which are merged (Table~\ref{tab:md}). In $\theta$ at 500~K these are 5 of the
10 snapshots of the $[100]$ edge dislocation along $[010]$ and all 10 of the $[110]$ screw dislocation, whose Burgers
vector is the largest ($|\bvec|=8.4$~\AA). Every other $\theta$ type is resolved in all 10 snapshots at both
temperatures. The 42 NiAl snapshots were also analyzed with the species-blind bcc reference, which can separate the
$\tfrac12[111]$ superpartials. The two superpartials of each superdislocation often share one
cluster, which returns the $\tfrac12[111]$ of each, and no snapshot was resolved: 26 were merged, 16 lost, and none
wrong. In the 16 lost snapshots, fewer than
eight clusters return a vector, but their boundary surfaces have at least eight handles in all, so superpartials of the
same sign share clusters and are read as single lines. In none of these snapshots are all eight superpartials separate
clusters.

\emph{Close to melting.} At 1500~K ($0.85\,T_\mathrm m$) the lines are resolved in 40 of 42 snapshots. At 1650~K
($0.93\,T_\mathrm m$) a quarter of the atoms are not classified as crystalline even in defect-free crystal (24\% in a defect-free
sample, 25\% in the samples with dislocations), and their clusters percolate. In all 42 snapshots the four lines then
lie in one box-spanning cluster of $1.5$--$2.3\times10^4$ atoms, which returns nothing but the vectors of the
dislocations it contains: its curves return $\pm\bvec$ (in two snapshots also $2\bvec$), or in one snapshot $\bvec$
alone, with all curves of the same sign.
That snapshot would read as lost, but its cluster holds all four lines, and we count it as merged. The same holds at
1700 and 1755~K (0.96 and $0.99\,T_\mathrm m$), where 28--35\% of the atoms are not classified as crystalline: in all 84 snapshots
the four lines lie in one cluster of $1.9$--$3.3\times10^4$ atoms, whose curves return $\pm\bvec$, in 14 snapshots
also $\pm2\bvec$, and nothing else. No sample melted during the run, since the potential energy and the crystalline fraction
stay constant. At these temperatures a sample analyzed whole does not resolve the lines individually, and a larger
sample would not help, since the thermal clusters percolate. The melting temperature of the potential,
$T_\mathrm m=1770$~K, was found by solid--liquid coexistence. A bcc box of $6.4\times10^4$ atoms ($20\times20\times80$
cells), half melted, was run at constant enthalpy and zero pressure for 600~ps, and its temperature settled at
1767--1770~K with both phases present.

\emph{Close-packed crystals.} Cu and Mg were analyzed with the table of cell topologies and neighbor agreement
(Supplementary Section~\ref{sec:si-degenerate}). Most stacking-fault atoms fail the checks of the host crystal, so
each dissociated Cu dislocation, with its two Shockley partials and the fault between them, forms one cluster
with the Burgers vector of the perfect dislocation, $\tfrac12[1\bar10]$. Every Cu snapshot gave four such clusters, with $\pm\tfrac12[1\bar10]$
with zero sum, and 94--98\% of the atoms passed the checks (97--98\% at 300~K, 94--96\% at 1200~K). Every Mg snapshot
gave four clusters with $\pm[0001]$, and 96--97\% of the atoms passed the checks. The other atoms form small clusters that return no
vector.

\subsection{Comparison with DXA}\label{sec:si-dxa}
\emph{bcc Fe.} We ran DXA (OVITO 3.17, default parameters, bcc input structure) on the larger bcc Fe edge quadrupole
and on the screw quadrupole of Table~\ref{tab:md}, 21 snapshots at each of 300, 900, 1200, 1500, and 1650~K. In every
snapshot DXA found segments with $\pm\tfrac12[111]$ and no other Burgers vector. Up to 1500~K the total length it
traced stayed close to that of the four lines, about 115~\AA\ for the edge and 160~\AA\ for the screw sample. At
1650~K ($0.93\,T_\mathrm m$) it fell to 68~\AA\ on average for the edge sample, so parts of those lines were not
traced, while the screw lines were traced almost fully (154~\AA). DXA classified 99.6--99.8\% of the atoms as bcc at 300~K,
89\% at 1200~K, and 54\% at 1650~K. At 1650~K, DXA thus still traces the four lines separately, though with gaps, while our
analysis joins them into one cluster spanning the box, which returns only the vectors of the lines it contains
(Section~\ref{sec:md}). Neither method returned a wrong vector.

\begin{figure}[!htbp]
\centering
\includegraphics[width=\textwidth]{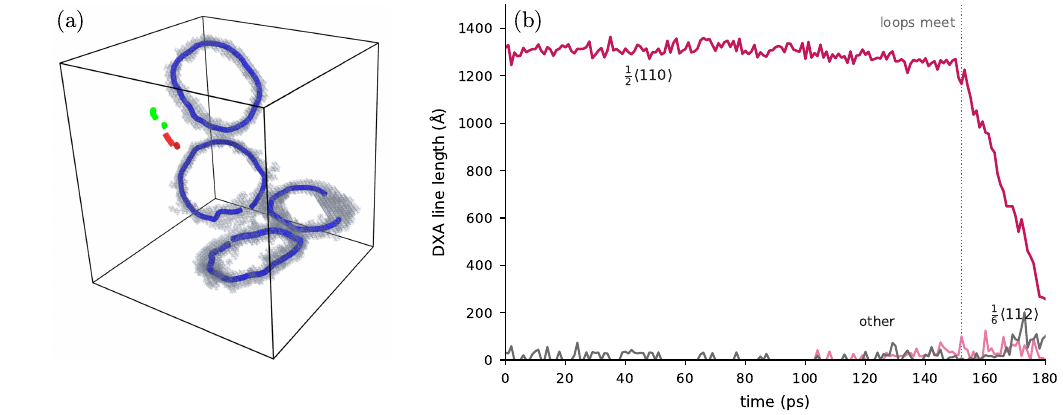}
\caption{\textbf{DXA on the Mg sublattice of C15 MgCu$_2$ during plastic flow.} (a) At 140~ps, before the loops
meet: atoms of our defect clusters with a Burgers vector (gray), and the lines DXA traces (blue,
$\tfrac12\langle110\rangle$; green, $\tfrac16\langle112\rangle$; red, other vectors). (b) Total length of DXA's
lines by Burgers vector: $\tfrac12\langle110\rangle$ (dark), $\tfrac16\langle112\rangle$ (light), and others (gray).
Dotted line: the loops first meet (152~ps).}
\label{fig:si-c15-dxa}
\end{figure}

\emph{C15 MgCu$_2$.} In C15 the Mg atoms form a diamond lattice, and LaCA identifies dislocations by running DXA on
this sublattice~\cite{laca2021}. We followed this procedure with OVITO's DXA, not with LaCA's own software. DXA (OVITO
3.17, default parameters, cubic-diamond input structure) was run on the Mg atoms alone in every snapshot of the C15
plastic-flow run of the main text, every 1~ps from 0 to 180~ps (181 snapshots). Burgers vectors are given in units of
the cubic cell, which C15 and its Mg sublattice share (Fig.~\ref{fig:si-c15-dxa}).

Up to 150~ps, DXA traced all four loops with their $\tfrac12\langle110\rangle$ vectors, the same loops and vectors as
our analysis, with 1{,}210--1{,}370~\AA\ of line in all. In some snapshots it also reported short segments with other
vectors, up to about 120~\AA\ in all. Our clusters at this stage are closed rings, one around each loop. DXA's lines
lie inside them but have gaps (Fig.~\ref{fig:si-c15-dxa}a).

The loops first meet at 152~ps. From then on DXA still finds three or four of the four $\tfrac12\langle110\rangle$
vectors in every snapshot, but the total length of line with these vectors shrinks to about 260~\AA\ at 180~ps, a fifth of its earlier
length, although the loops have grown. DXA also reports segments with $\tfrac16\langle112\rangle$, from about 100~ps on, the Burgers vector of synchro-Shockley partials in Laves
phases~\cite{xie2023synchroshockley}, which may be real partials that our analysis does not resolve. It also reports
segments with vectors that are not lattice vectors of the crystal, such as $\tfrac14\langle111\rangle$, $\tfrac14\langle113\rangle$, $\tfrac1{12}\langle511\rangle$,
and $\tfrac1{12}\langle111\rangle$, some of them from the first snapshots on. Over the same snapshots, every branch in our analysis
returns one of the four seeded vectors (main text).

A dislocation line cannot end inside crystal, so open ends mark lines that were not fully traced or segments that are
not dislocations. Our analysis cannot produce them (Discussion).

\subsection{Run time and memory}\label{sec:si-runtime}
Each sample is a periodic molecular-dynamics snapshot tiled $n\times n\times n$ times: bcc Fe with prismatic loops at
900~K (45{,}815 atoms per tile), screw and edge quadrupoles in bcc Fe, dissociated edges in fcc Cu, and $[0001]$ screws
in hcp Mg, all at 300~K, the $\theta$-Al$_2$Cu junction at 200~ps, and the C15 network at 180~ps. Every case ran alone on a
whole node, one after another (on the Platinum 8558 node, 46 of its 96 physical cores, the most its queue allows). Times run from the positions in memory to the Burgers vectors of all clusters, for ours and for DXA (OVITO 3.17, default
parameters, with the bcc, fcc, or hcp input structure), and exclude reading the file. Our analysis runs as a Python
process and C++ child processes, so its peak memory is the largest sum, over its processes, of their proportional set
sizes, sampled every 0.1~s (for the 79-million-atom sample, their resident memory, larger by under
1\%). For DXA it is the peak resident memory. The four node types are two 28-core Xeon Gold 6348
processors (2.6~GHz), one 32-core Xeon Gold 6338 (2.0~GHz), two 18-core Xeon Gold 6240 (2.6~GHz), and two 48-core Xeon
Platinum 8558.

Across the four node types, times differ by up to a factor of 1.7 (Table~\ref{tab:si-runtime}), and the ratios to DXA
given in Methods hold on every one. On a Gold 6348 node, from 1 to
16 threads our analysis runs 4.8--6.8 times faster, and DXA 1.5 times (Table~\ref{tab:si-threads}). The gain is
similar on the Gold 6338 and Gold 6240 nodes and smaller on the Platinum 8558 node, 2.7--4.2 times
(Table~\ref{tab:si-runtime}). Beyond 16 threads the tessellation, computed with the multithreaded voro++
containers~\cite{lu2023voromt}, became slower: for 9.9 million atoms of bcc Fe it took 14~s on 16 threads and 64~s on
56, while the other stages together became slightly faster.

\begin{table}[!htbp]
\centering\footnotesize\setlength{\tabcolsep}{3.5pt}
\begin{tabular}{lrrrrrrrrr}
\toprule
 & atoms & \multicolumn{2}{c}{Gold 6348} & \multicolumn{2}{c}{Gold 6338} & \multicolumn{2}{c}{Gold 6240} &
\multicolumn{2}{c}{Platinum 8558}\\
sample & ($10^6$) & ours & DXA & ours & DXA & ours & DXA & ours & DXA\\
\midrule
bcc Fe, prismatic loops & 1.24 & 7.8 & 15.1 & 7.9 & 19.2 & 9.7 & 16.8 & 11.7 & 17.2\\
 & 2.93 & 18.8 & 34.3 & 19.3 & 43.3 & 22.9 & 38.3 & 25.3 & 35.1\\
 & 9.90 & 66.5 & 113.4 & 65.2 & 139.8 & 80.6 & 124.0 & 67.0 & 94.5\\
 & 23.46 & 163.5 & 265.0 & 162.7 & 333.8 & 196.4 & 292.0 & 138.9 & 218.0\\
 & 45.82 & 328.2 & 519.9 & 318.6 & 643.5 & 410.3 & 622.0 & 270.8 & 418.3\\
 & 79.17 & 591.8 & 915.3 & 573.9 & 1108.2 & 809.4 & 1017.3 & 483.4 & 738.8\\
bcc Fe, screw quadrupole & 1.55 & 9.7 & 18.6 & 9.7 & 22.8 & 11.5 & 20.4 & 12.8 & 21.8\\
bcc Fe, edge quadrupole & 2.12 & 14.1 & 25.7 & 14.7 & 30.9 & 16.9 & 28.2 & 17.9 & 21.5\\
$\theta$-Al$_2$Cu, junction & 2.27 & 14.7 & -- & 15.0 & -- & 19.4 & -- & 18.9 & --\\
 & 18.20 & 129.0 & -- & 139.8 & -- & 162.5 & -- & 114.7 & --\\
C15 MgCu$_2$, network & 1.22 & 15.0 & -- & 15.5 & -- & 17.8 & -- & 20.1 & --\\
 & 9.73 & 264.1 & -- & 280.3 & -- & 312.0 & -- & 219.8 & --\\
fcc Cu, dissociated edges & 1.94 & 17.9 & 27.1 & 18.6 & 32.8 & 20.8 & 29.2 & 22.2 & 28.9\\
 & 15.53 & 145.6 & 198.8 & 146.6 & 246.2 & 169.1 & 218.6 & 123.8 & 166.1\\
hcp Mg, $[0001]$ screws & 1.32 & 18.1 & 18.0 & 18.9 & 21.8 & 20.7 & 19.9 & 23.2 & 21.2\\
 & 10.58 & 150.6 & 135.1 & 157.2 & 162.2 & 172.0 & 152.5 & 133.0 & 117.6\\
\midrule
\multicolumn{10}{l}{one thread}\\
bcc Fe, prismatic loops & 1.24 & 39.6 & 22.4 & 43.7 & 27.0 & 44.8 & 24.6 & 31.6 & 27.3\\
$\theta$-Al$_2$Cu, junction & 2.27 & 101.4 & -- & 107.6 & -- & 114.8 & -- & 78.6 & --\\
fcc Cu, dissociated edges & 1.94 & 89.7 & 41.5 & 96.2 & 48.0 & 96.3 & 44.5 & 77.9 & 33.4\\
\bottomrule
\end{tabular}
\caption{Run time (s) for one snapshot on four node types, with 16 threads and with one thread.}
\label{tab:si-runtime}
\end{table}

\begin{table}[!htbp]
\centering\footnotesize
\begin{tabular}{llrrrrrrr}
\toprule
 & & \multicolumn{7}{c}{threads}\\
 & & 1 & 2 & 4 & 8 & 16 & 28 & 56\\
\midrule
bcc Fe & ours & 321.4 & 189.2 & 120.0 & 83.5 & 67.6 & 76.4 & 117.4\\
 & DXA & 174.4 & 141.4 & 125.2 & 120.0 & 113.1 & 111.5 & 110.1\\
fcc Cu & ours & 89.7 & 52.2 & 33.6 & 23.1 & 17.9 & 15.7 & 14.2\\
 & DXA & 41.4 & 33.5 & 29.5 & 27.9 & 26.8 & 26.2 & 26.1\\
$\theta$-Al$_2$Cu & ours & 101.6 & 53.7 & 33.0 & 20.3 & 14.9 & 17.7 & 21.8\\
\bottomrule
\end{tabular}
\caption{Run time (s) against the number of threads on a second Gold 6348 node, so that times at 16 threads differ
slightly from Table~\ref{tab:si-runtime}, for bcc Fe with prismatic loops
(9.90 million atoms), fcc Cu with dissociated edges (1.94 million), and the $\theta$-Al$_2$Cu junction (2.27 million).}
\label{tab:si-threads}
\end{table}

\section*{Supplementary Movies}\addcontentsline{toc}{section}{Supplementary Movies}
\noindent The movies can be viewed and downloaded at \url{https://www.vorotop.org/burgers-movies.html}.

\medskip\noindent\textbf{Supplementary Movie 1. Burgers vectors from Voronoi topology in two dimensions.} A
two-dimensional Lennard-Jones crystal of 440 atoms in a periodic box, heated to $k_\mathrm{B}T/\epsilon =
0.64$ and held. Atoms are colored by the number of sides of their Voronoi cells, as in Ref.~\cite{lazar2024twod}.
At the first bond flip the movie pauses and builds, step by step, its defect cluster, boundary, shell, and
circuit. Each step of the circuit is drawn in the ideal crystal on the right, and the steps close. The crystal is then
slowly sheared, and a bond flip splits into two dislocations. To show the pair close together, the run is continued
from 0.15 time units after the split with velocities redrawn at the same temperature, and the movie shows one of 1,200
such continuations, in which one dislocation has merged with a thermal fluctuation. The circuits of the two clusters
give $\bvec$ and $-\bvec$, and a circuit around both gives zero.

\medskip\noindent\textbf{Supplementary Movie 2. C15 MgCu$_2$ during plastic flow at 300~K.} The sample of
Fig.~\ref{fig:showcase}a, every 1~ps from the start of straining to 180~ps. Atoms of the defect clusters with a nonzero
Burgers vector are colored by the vector of their branch, as in Fig.~\ref{fig:showcase}. Each branch is drawn whole
across the periodic boundaries unless it is comparable in size to the box.

\medskip\noindent\textbf{Supplementary Movie 3. $\theta$-Al$_2$Cu during plastic flow at 300~K.} The sample of
Fig.~\ref{fig:showcase}b, every 1~ps from the start of straining to 200~ps, drawn as in Supplementary Movie~2.

\medskip\noindent\textbf{Supplementary Movie 4. Crossing loops in $\theta$-Al$_2$Cu at 300~K, seen from all sides.}
The crossing loops of Fig.~\ref{fig:junction}, every 1~ps from 0 to 359~ps, while the camera circles the sample once,
one degree per snapshot. Atoms of the defect cluster are colored by the Burgers vector of their branch, as in
Fig.~\ref{fig:junction}, and crystal atoms are not drawn. The $[100]$ segment is present in every snapshot.

\medskip\noindent\textbf{Supplementary Movie 5. Crossing loops in $\theta$-Al$_2$Cu heated in steps.} The sample of
Fig.~\ref{fig:junction}, held at 150~K for 50~ps and then heated to 300, 450, and 600~K, with 25~ps ramps
and 50~ps holds, every 0.5~ps (every 0.25~ps during heating). The view and colors are those of
Fig.~\ref{fig:junction}: atoms of the defect cluster are colored by the Burgers vector of their branch, and one fixed
set of crystal atoms on the two glide planes by cell topology. The $[100]$ segment is present in every
snapshot. The dark cells, matching neither ideal cell, are thermal fluctuations and return no Burgers vector.

\medskip\noindent\textbf{Supplementary Movie 6. Two loops in $\theta$-Al$_2$Cu that join along a $[100]$ segment at 300~K.} The sample in
which each loop stops 8~\AA\ short of the line where the two glide planes meet, from 0 to 200~ps. The dislocations are
drawn as closed curves inside their defect cluster, smoothed for display and colored by Burgers vector.
\end{document}